\documentclass[twocolumn]{aastex63}
\usepackage{apjfonts}
\usepackage{epsfig}
\usepackage{epstopdf}
\usepackage{natbib}
\usepackage{amssymb,latexsym}
\usepackage{graphicx}
\usepackage{appendix}
\usepackage{amsmath}

\usepackage[normalem]{ulem}
\newcommand{\be}{\begin{equation}}
\newcommand{\ee}{\end{equation}}

\newcommand{\eduardom}{$\rm{redMaPPer}$}
\newcommand{\redmapper}{$\rm{redMaPPer}$}

\shortauthors{To et al}
\shorttitle{RedMaPPer CLFs}

\begin{document}
\title{RedMaPPer: Evolution and Mass Dependence of the Conditional Luminosity Functions of Red Galaxies in Galaxy Clusters}
\correspondingauthor{Chun-Hao To}
\email{chto@stanford.edu}
\author[0000-0001-7836-2261]{Chun-Hao To}
\affiliation{Kavli Institute for Particle Astrophysics and Cosmology}
\affiliation{Physics Department, Stanford University, Stanford, CA, 94305}
\affiliation{SLAC National Accelerator Laboratory, Menlo Park, CA, 94025}
\author{Rachel M. Reddick}
\affiliation{Kavli Institute for Particle Astrophysics and Cosmology}
\affiliation{Physics Department, Stanford University, Stanford, CA, 94305}
\affiliation{SLAC National Accelerator Laboratory, Menlo Park, CA, 94025}
\author{Eduardo Rozo}
\affiliation{Department of Physics, University of Arizona, Tucson, AZ 85721, USA}
\author{Eli Rykoff}
\affiliation{Kavli Institute for Particle Astrophysics and Cosmology}
\affiliation{SLAC National Accelerator Laboratory, Menlo Park, CA, 94025}

\author[0000-0003-2229-011X]{Risa H. Wechsler}
\affiliation{Kavli Institute for Particle Astrophysics and Cosmology}
\affiliation{Physics Department, Stanford University, Stanford, CA, 94305}
\affiliation{SLAC National Accelerator Laboratory, Menlo Park, CA, 94025}

\begin{abstract}
We characterize the luminosity distribution, halo mass dependence, and redshift evolution of red galaxies in galaxy clusters using the SDSS Data Release 8 \redmapper{} cluster sample. We propose a simple prescription for the relationship between the luminosity of both red central and red satellite galaxies and the mass of their host halos, and show that this model is well-fit by the data.
Using a larger galaxy cluster sample than previously employed in the literature, we find that the luminosities of red central galaxies scale as $\langle \log L \rangle \propto A_L \log (M_{200b})$, with $A_L=0.39\pm0.04$, and that the scatter of the red central--galaxy luminosity at fixed $M_{200b}$ ( $\sigma_{\log L|M}$)
 is $0.23 ^{+0.05}_{-0.04}$ dex, with the error bar including systematics due to miscentering of the cluster finder, photometry, and photometric redshift estimation. 
Our data prefers a positive correlation between the luminosity of red central galaxies and the observed richness of clusters at a fixed halo mass, with an effective correlation coefficient $d_{\rm{eff}}=0.36^{+0.17}_{-0.16}$. The characteristic luminosity of red satellites becomes dimmer from $z=0.3$ to $z=0.1$ by $\sim 20\%$ after accounting for passive evolution. We estimate the fraction of galaxy clusters where the brightest red galaxy is not the central to be $P_{\rm{BNC}} \sim 20\%$.
We discuss implications of these findings in the context of galaxy evolution and the galaxy--halo connection. 
\end{abstract}
\keywords{galaxy evolution --- large-scale structure of universe --- galaxy dark matter halos --- galaxy clusters}

%\maketitle

\section{Introduction}
Galaxy clusters form from the highest density peaks of the matter density field, making them interesting objects to study both cosmology and astrophysics. Cosmologically, the abundance of clusters is sensitive to structure formation, and the redshift evolution of the abundance function is a powerful probe of the dark energy equation of state. Astrophysically, the crowded environments of galaxy clusters provide an important laboratory for studying galaxy formation and evolution. Galaxies that fall into clusters make a distinct transition from star forming to quenched as their gas and dark matter are stripped away (e.g., \citealt{2008MNRAS.387...79V, 2010ApJ...721..193P, Wu2013, WTC2011}). Some of the galaxies are even entirely destroyed, with their luminous matter dispersed into the intra-cluster light or accreted onto the central galaxy of the cluster \citep{CWK2007, WBZ2012, Zhang18}. A close examination of the luminosity distribution of galaxies in galaxy clusters and their redshift evolution enables us to test how these processes occur. 

In the current paradigm of structure formation, all galaxies are assumed to form inside dark matter halos. Therefore, it is natural to assume that the properties of galaxies are connected to the properties of the dark matter halos they live in. As summarized in \cite{RisaAwesomepaper}, various models have been proposed to describe the connection between galaxies and dark matter halos. These models range from models in which one directly simulates or parameterizes the physics of galaxy formation, to empirical models that assume an {\em ad hoc} functional form for the galaxy--halo connection that is constrained from data. Here we focus on a purely empirical approach related to the Halo Occupation Distribution (HOD), which associates the distribution in the number of galaxies of a given property to the mass of their host halos. The conditional luminosity function (CLF) is a more detailed version of the HOD model that parameterizes the galaxy occupancy of halos as a function of galaxy luminosity and/or stellar mass and/or color. There are now many empirical constraints on the CLF \citep{LMS2004, YMB2008, YMB2009, Hansen09, Cacciato2013, Zhang2017}. Precise and accurate measurement of the CLF model parameters can shed light on various astrophysical effects, such as the strength of AGN feedback \citep{Kravstov2018} and the redshift evolution of cluster galaxies  \citep{Zhang2017}. This measurement also provides a direct constraint on the galaxy--halo connection, facilitating cosmological studies that use galaxies as tracers of the dark matter density field. These studies also enable us to constrain the scatter of luminosity--halo mass relations \citep{YMB2009, Kravstov2018}, and to predict the rate at which the central galaxy of a halo is {\it not} the brightest galaxy within that halo \citep{Ski2011, Lange2018}. 

The parameters of CLF models have been inferred from a combination of galaxy clustering, galaxy--galaxy lensing, and galaxy luminosity distribution measurements \citep[e.g.][]{Cacciato2013}, as well as from direct measurement from groups and cluster catalogs \citep{LMS2004, YMB2008, YMB2009, Hansen09}. Each of these methods presents its own set of systematics and limitations. Studies based on galaxy clustering are mostly sensitive to low--mass systems, and therefore fail to provide a tight constraint on the CLF of massive systems. Direct measurements using cluster or group catalogs must rely on an accurate calibration of the observable--halo mass relation, and require proper modeling of possible correlation between observables; these correlations have generally not been included in previous work.  Finally, direct measurements are also sensitive to systematics associated with cluster finding. 

In this paper, we measure the red galaxy conditional luminosity function using the Sloan Digital Sky Survey (SDSS) \citep{SDSS2011} \redmapper{} cluster catalog \citep{Redmapper1}.  Relying on the cluster mass calibration of \citet{simetetal17} used in the cosmological analysis of this sample presented in \cite{SDSS_cluster_cosmology}, we fit for the mass dependence of the red galaxy CLF. We marginalize over the possible correlation between observables, and account for what we expect are the primary systematics in this dataset, namely photometric biases, centering errors in the \redmapper{} catalog, and cluster photometric redshift uncertainty. Importantly \citep[see e.g.][]{covariance}, our fits rely on the full covariance matrix across across bins of luminosity in our data vector.   We constrain how the CLF of massive halos depends on halo mass and redshift, and use our results to predict the rate at which the brightest galaxy in a halo is not the central galaxy.

The paper is laid out as follows. In Section \ref{sec-data} we present the data sets used in this analysis, including a brief overview of the \redmapper\ algorithm (Section~\ref{subsec-redm}), and the calibration of cluster membership using SDSS and Galaxy and Mass Assembly (GAMA) spectroscopy (Section~\ref{subsec-prob}). We describe an empirical correction we apply to bright SDSS galaxies in Section \ref{subsec-photometry}.  We explain our approach for obtaining the CLF from the \redmapper\ data, and describe our estimate of the covariance  matrix in Section~\ref{sec-clf}. In Section \ref{sec:modeling}, we present our model of the conditional luminosity function. We address possible systematics in Section \ref{sec-sys}. We summarize our key results in section \ref{sec-ev}, and discuss their implications in Section \ref{sec-discuss}. In particular, we investigate the relationship between the cluster central galaxy and the brightest cluster galaxy in Section~\ref{sec-pbcg}.  Section~\ref{sec-summary} summarizes our conclusions.

Throughout this paper, we assume that $H_0$ is $68.2\rm{km~s^{-1}~Mpc^{-1}}$ and set $h\equiv H_0/(100~\rm{km~s^{-1}~Mpc^{-1}})$ to $0.682$.  Where necessary, we assume a flat $\Lambda$CDM cosmology with $\Omega_m=0.301$, as in the DES Y1 cosmology result (\citealt{DESY1}). Through our the paper, we define halo mass as $M_{200b}$ where $M_{200b}=\frac{4}{3}\pi R_{200b}^3 200\bar \rho_{\rm m}$, and $R_{200}$ is the radius at which the averaged enclosed density of the halo is $200$ times larger than the mean density of the universe $\bar \rho_{\rm m}$.

\section{Data}\label{sec-data}

The analysis is performed on SDSS DR8 data \citep{Redmapper1}, which covers approximately 10405~$\rm{deg}^2$ with $i$-band depth of $\sim 20.9$. When calculating the CLF, we use absolute magnitudes derived from the {\sc kcorrect} code of \cite{BlRo2007}, $k$-corrected to $z=0.3$ with a fixed red galaxy template.  We then convert this absolute magnitude into solar luminosities, and we use this $i$-band solar luminosity for all calculations of the CLF.  We do not correct galaxy luminosities for passive evolution when estimating the CLF.

We use data from the Galaxy and Mass Assembly survey \citep[GAMA][]{Dri2011} to calibrate the likelihood of a photometric galaxy being a spectroscopic cluster member, as discussed in Section~\ref{subsec-prob} and described in detail in a related paper \citep{Rozo2014}.

\subsection{\eduardom{}}\label{subsec-redm}

Cluster--finding is performed using the red-sequence based cluster finder \eduardom{} \citep{Ryk2013,RoRy2013}, which identifies galaxy clusters as overdensities of red-sequence galaxies. Importantly, \redmapper{} includes a probabilistic assignment of galaxy membership, allowing straightforward incorporation of uncertainty regarding whether a galaxy is or is not a red galaxy in a particular cluster. We further discuss these probabilities in Section~\ref{subsec-prob}. However, it is important to note that this catalog only includes red galaxies; thus, the CLFs presented in this paper are for red galaxies only. As shown in \cite{Wet13}, more than $80\%$ of the centrals and $50\%$ of the satellites living in the mass and luminosity range considered in this paper are quenched; thus, the results presented here represent the properties of the majority of the cluster galaxy population, but the details will clearly differ compared to the full cluster galaxy population. 

In this paper, we consider the \redmapper\ v5.10 cluster catalog derived from the SDSS DR8 data set \citep{Rozo2014}.  The depth of DR8 allows us to select a volume-limited cluster catalog out to a redshift $z\leq 0.3$. The redshift limit corresponds to the redshift at which luminosity threshold
of $0.2L_*$ used by \redmapper\ crosses the survey depth.  Here, $L_*$ is the passively evolving characteristic luminosity of cluster galaxies assumed by \eduardom{} in its filtering process. The final sample contains 7016 clusters with $\lambda>20$ in the redshift range of $0.1<z<0.3$.

\subsection{Probabilities with \redmapper\ }\label{subsec-prob}

For every galaxy in the vicinity of a galaxy cluster, the \redmapper\ algorithm estimates the probability that the galaxy is a cluster member
on the red-sequence. Comparison of the photometric probabilities with spectroscopic data from SDSS \citep{SDSS2011} and GAMA \citep{Dri2011} led \citet{Rozo2015} to derive small corrections to the original membership probabilities.  The analysis in this work relies on these improved membership probabilities. 

In addition to providing galaxy membership probabilities, \redmapper\ also assigns cluster centers in a probabilistic fashion.   While most clusters have a single bright galaxy clearly located at the cluster center, for others it is not possible to unambiguously identify a unique central galaxy. Consequently, \eduardom{} provides a list of possible central galaxies, each tagged with the probability of it being the central galaxy of the cluster.  Many clusters have more than a single high-probability center: about 63\% (34 \%) of clusters in our DR8 sample with $\lambda>20$ have at least two galaxies with a greater than 1\% (10\%) probability of being the central galaxy. This necessitates approaching the problem of cluster membership and cluster centering in a probabilistic way, especially when investigating the properties of central galaxies. In our fiducial analysis, we assume that the \redmapper{} centering probability estimates are correct \citep{Rozo2014}, and model the resulting ensemble properties in order to constrain the behavior of central galaxies in massive cluster halos. We discuss how this assumption affects our results in Section \ref{sec-sys}, and discuss how to incorporate this effect into our error budget.

\subsection{Photometry with \redmapper\ }\label{subsec-photometry}

\begin{figure}
\centering
\includegraphics[width=0.5\textwidth]{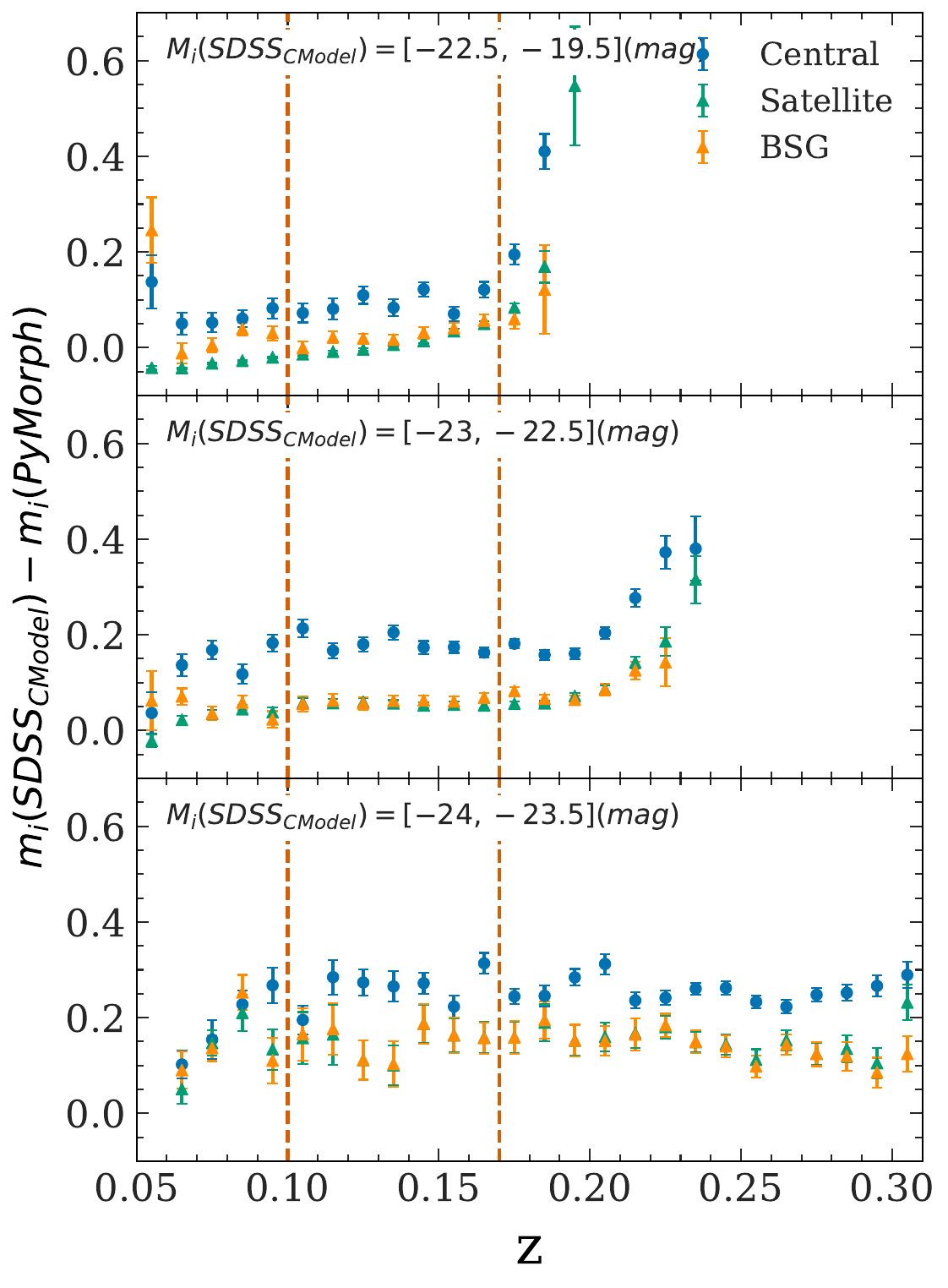}\hspace{-0.05\textwidth}
\caption{Difference between SDSS \textbf{CModel} and \textbf{PyMorph} photometry as a function of redshift in three different magnitude bins. Blue symbols represent red central galaxies, green symbols represent red satellites, and orange symbols represent the brightest red satellite galaxies  (BSGs) in each cluster. Again, we find that the
correction for the brightest red satellite galaxies are fully consistent with
correction for other  red satellite galaxies, providing further evidence for
the fidelity of central galaxy identification. We find no redshift dependence in the photometry difference for samples t isn the brightest magnitude bins (bottom panel), and mild redshift dependence for fainter samples (top two panels). Since high redshift samples are prone to selection effects and correcting those effects is beyond the scope of this paper, we adopt an empirical redshift cut at $z=0.17$ while calculating the empirical photometry correction.}
\label{fig-bernardi_redmapper_Z}
\end{figure}

Galaxy luminosities in the SDSS \redmapper\ catalog were calculated based on the SDSS \textbf{CModel} magnitudes, which are known to underestimate galaxy brightness for massive galaxies \citep{Ber13}. This bias depends on galaxy luminosity and type \citep{Ber17} and thus has a large impact on the inference of the galaxy luminosity function.  
To account for this, \cite{Meert2015} performed an improved photometric measurement (\textbf{PyMorph} magnitude hereafter) on SDSS DR7 spectroscopic targets. We rely on the \cite{Meert2015} measurement to develop an empirical correction for the SDSS photometry of each \redmapper{} galaxy. 

\begin{figure}
\centering
\includegraphics[width=0.5\textwidth]{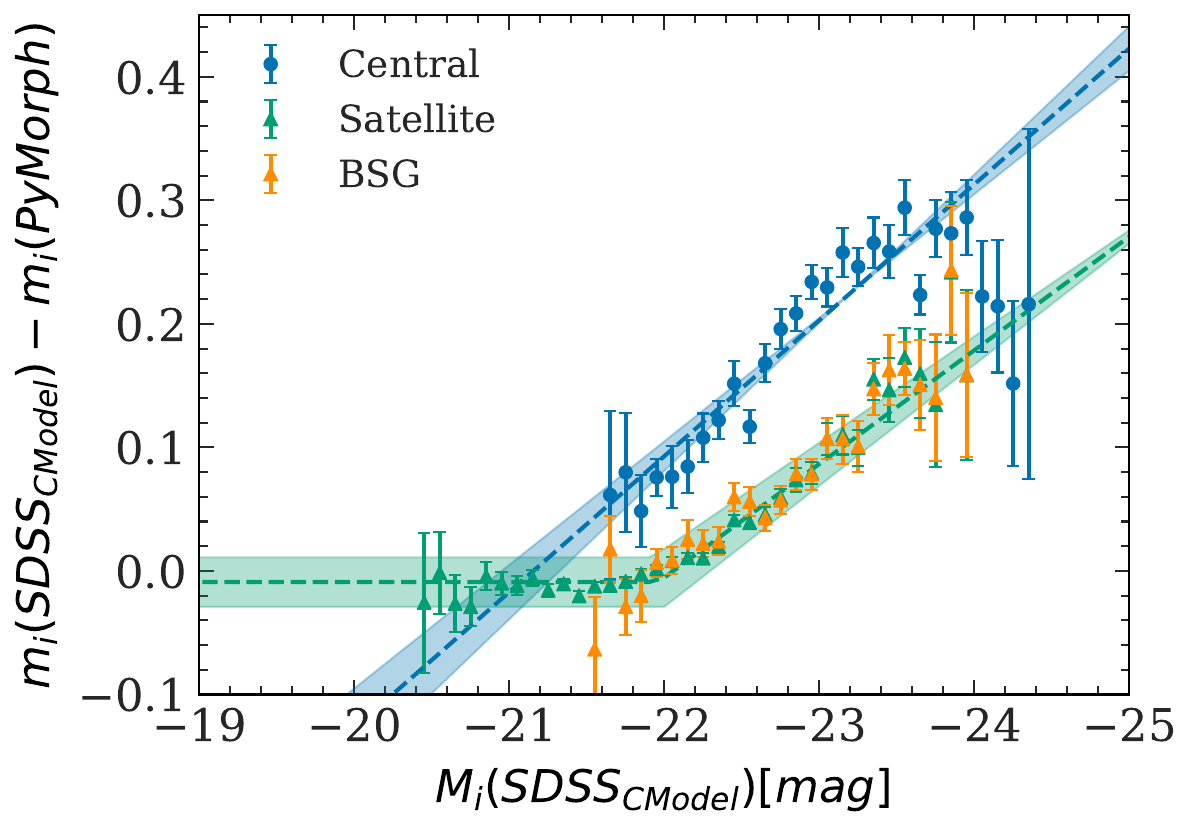}\hspace{-0.05\textwidth}
\caption{Difference between SDSS \textbf{CModel} and \textbf{PyMorph} photometry as a function of SDSS \textbf{CModel} for clusters with $\lambda >20$ and $0.1<z<0.17$. Blue symbols represent  red central galaxies, green symbols represent  red satellites, and orange symbols represent the brightest  red satellite galaxies (BSGs) in each cluster. We find that the correction for the brightest  red satellite galaxies are fully consistent with correction for other satellite galaxies, providing further evidence for the fidelity of central galaxy identification. The blue and green lines show the empirical correction we obtain by fitting equation \ref{eqn:Central} and \ref{eqn:Sat} to the points, using the parameters summarized in Table \ref{tab-correct-params}. This correction is applied to the full samples in the analysis.}
\label{fig-bernardi_redmapper_M}
\end{figure}

The empirical correction is calculated as follows. First, we select galaxies that have good measurements of total $i$ band magnitude (without \textbf{Flag} above 20) in the catalog described in \citet{Meert2015} and \citet{Meert2016}. For each galaxy we take the "best model" magnitude (\textbf{PyMorph}) described in \cite{Meert2015}, which is estimated from a combination of Sersic and Sersic--Exp profile fits. Second, we cross-match the above catalog to \redmapper{} member galaxies by matching galaxies within $3"$ and redshift differences within $0.03$. Here, we assume that member galaxies have the same redshifts as reported in \redmapper{} catalog. Third, we compute the difference between the SDSS \textbf{CModel} magnitude and \textbf{PyMorph} magnitude.  We look for any redshift dependence of this magnitude difference in three absolute magnitude bins (Figure \ref{fig-bernardi_redmapper_Z}). We notice that at the faintest absolute magnitude bins, the difference between \textbf{CModel} and \textbf{PyMorph} becomes significant at $z>0.17$. The samples with \textbf{PyMorph} measurements are selected to be brighter than $m_r=17.77$, because this is the lower limit for completeness of the SDSS Spectroscopic Survey \citep{Meert2015}. The magnitude cut $m_r=17.77$ corresponds to $M_r=-21.81$ at $z=0.17$, which lies within the faintest magnitude bin in Figure \ref{fig-bernardi_redmapper_Z}.  Therefore, the redshift dependence we find in Figure \ref{fig-bernardi_redmapper_Z} is likely due to the incompleteness of the sample. To account for this selection effect we adopt an upper redshift cut at $z=0.17$  and a lower redshift cut $z=0.1$ (to match the \redmapper\ selection).   After this step, we obtain $18430$ matched galaxies, including $1516$ galaxies that are the most probable central galaxies in their host clusters.  We compute the mean difference between the \textbf{CModel} and \textbf{PyMorph} magnitude as a function of \textbf{CModel} magnitude for both  red central and  red satellite cluster galaxies. As shown in Figure \ref{fig-bernardi_redmapper_M}, this difference depends on whether a galaxy is a central or a satellite. The result is consistent with what \cite{Ber17} found. We further check the difference for brightest  red satellite galaxies and find that they are consistent with the full population of  red satellite galaxies. This gives further evidence for the fidelity of central galaxy identification.  Finally, we fit empirical central and satellite correction models to the observed magnitude differences. The correction models take the following form, 
\begin{align}
\label{eqn:Central}
Cent&ral: \nonumber \\
&\Delta m = A_{\rm{c}}(M_{\rm{CModel}}+22.5)+B_{\rm{c}} \\
\label{eqn:Sat}
Sate&llite:&\nonumber \\
&\Delta m = A_{s}(M_{\rm{CModel}}-B_{s}),\ if\ M_{\rm{CModel}} < B_{s} \nonumber\\
&\ \ \ \ \ =C_{s},\ \rm{if}\ M_{\rm{CModel}}> B_{s}, 
\end{align}
where $\Delta m$ is $m_{\textbf{CModel}}-m_{\textbf{PyMorph}}$

The best-fit parameters for the model are listed in Table \ref{tab-correct-params}. We apply this correction to every \redmapper{} galaxy based on the combination of central and satellite correction models weighting by the probability of a galaxy being a central.  

\begin{table*}
	\center
	\caption{Values of Empirical Photometry Correction Parameters}
	\begin{tabular}{l c c c }
	\hline \hline
	Parameters	&	Values & Equation reference(s) & Description \\
	\hline
	$A_{c}$	&		$-0.110^{+0.010}_{-0.010}$&	\ref{eqn:Central}	& Slope of Correction \\
    $B_{c}$	&		$0.148^{+0.007}_{-0.007}$&	\ref{eqn:Central}	&  Normalization of Correction \\
    $A_{s}$	&		$-0.092^{+0.006}_{-0.006}$&	\ref{eqn:Sat}	& Slope of Correction \\
    $B_{s}$	&		$-21.96^{+0.04}_{-0.04}$&	\ref{eqn:Sat}	& Cutoff of Correction\\
    $C_{s}$	&		$-0.009^{+0.002}_{-0.002}$& \ref{eqn:Sat}	& Constant after cutoff\\
	\end{tabular}
	\label{tab-correct-params}
\end{table*}

We discuss the impact of this photometric correction on our result in Section \ref{sec-sys}.

\section{Measurement of the Conditional Luminosity Function}
\label{sec-clf}

We measure the Conditional Luminosity Function (CLF) $\Phi$ in bins of redshift and cluster richness $\lambda$.  The CLF $\Phi$ is further separated into two parts: the satellite CLF, $\Phi_s$, and the central CLF, $\Phi_c$.  For the central CLF, we weigh our sum using the centering probabilities $p_{\rm{cen}}$, the probability that a given galaxy is a central galaxy. For the satellites, we use the membership probabilities $p_{\rm{mem}}$ multiplied by $1-p_{\rm{cen}}$ to account for the probability that a given galaxy is a satellite member galaxy, and not the central.

Expressing the CLF for a single bin with all clusters in $\lambda_{\rm{min}}<\lambda<\lambda_{\rm{max}}$ and $z_{\rm{min}}<z<z_{\rm{max}}$, we write:

\begin{align}
    \label{eq:measurement}
	\Phi_c(L) &= \frac{\sum_{i\in \rm{clusters}} \sum_{j\in \rm{galaxies~in~i}} p_{\rm{cen},j} }{ N_{\rm{cl}} \Delta \log L} \\
	\Phi_s(L) &= \frac{\sum_{i\in \rm{clusters}} \sum_{j\in \rm{galaxies~in~i}} p_{\rm{mem},j} (1-p_{\rm{cen},j}) }{ N_{\rm{cl}} \Delta \log L}
\end{align}

Here, $N_{\rm{cl}}$ is the number of clusters with richness and redshift in the bins being considered. We measure the conditional luminosity function in four evenly spaced redshift bins ranging from $z=0.1$ to $z=0.3$, and five richness bins, $\lambda = [20,25], [25,30], [30,40], [40,60], [60,100]$.

\begin{figure*}
\includegraphics[width=1.0\textwidth]{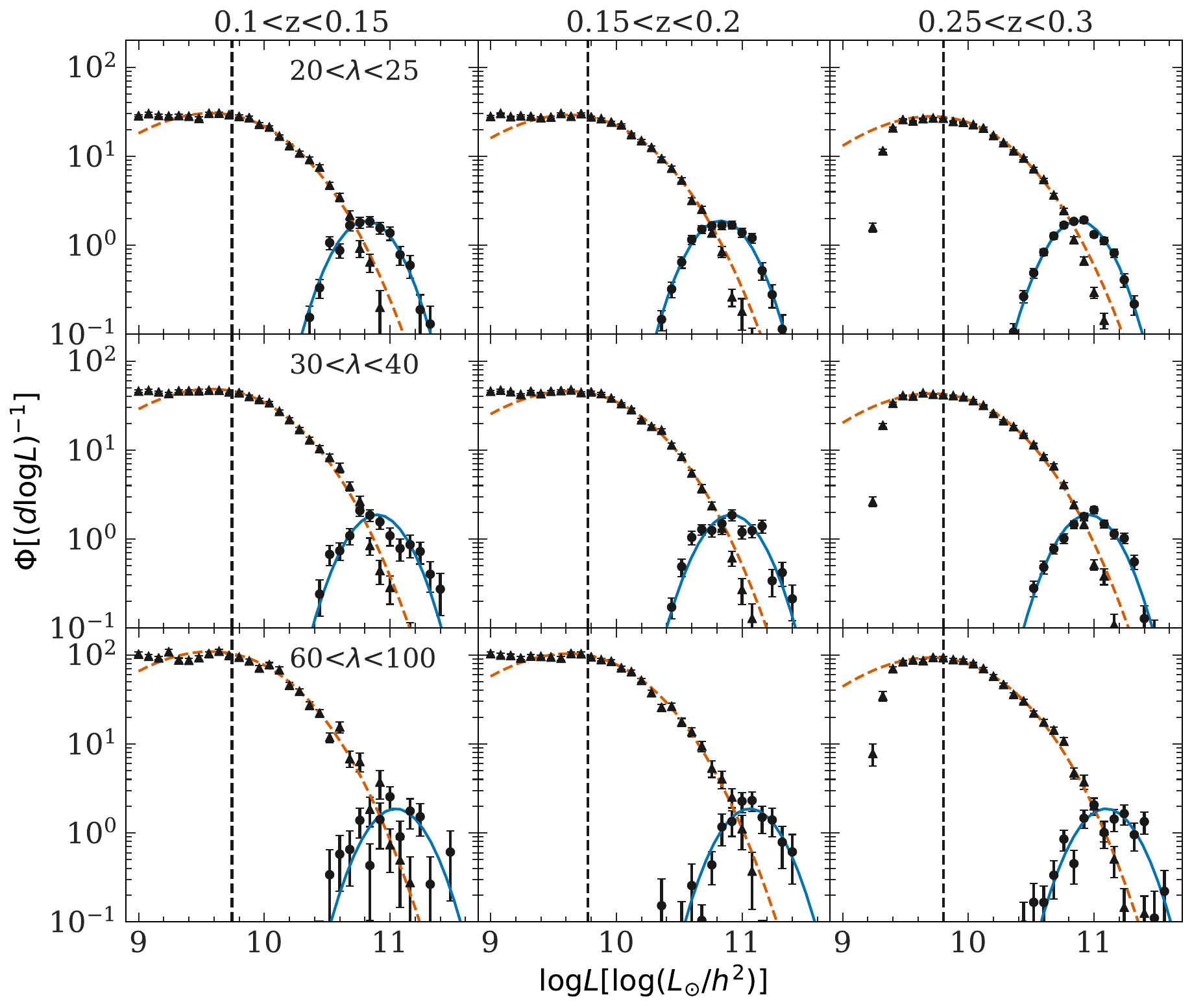}
\caption{CLF as a function of redshift $z$ and richness $\lambda$.  Richness increases from top to bottom, and redshift increases from left to right.  Solid black points and black triangles with error bars represent the measured luminosity function of red centrals and red satellites, respectively.  
The solid blue line is the fit we obtain for the  red central part of the model, and the dashed red line is the fit for the  red satellites.
The dashed, vertical black line shows the value of $0.317L^{*}$, corresponding to $0.2 L^*_{redm} + 0.2$dex, which is the lower bound of the data being considered in the model. 
Not all bins used in the fitting are shown; bin edges in redshift are $z=[0.1,0.15,0.2,0.25,0.3]$ and the bin edges in richness are $\lambda=[20,25,30,40,60,100]$.}
\label{fig-dr8clf}
\end{figure*}

The resulting CLFs are shown in Figure~\ref{fig-dr8clf} along with the fitted model described in Section \ref{sec:modeling}.

To determine the covariance matrix of the central CLF data, we consider both a theory-based covariance matrix and a jackknife estimate.  In Appendix~\ref{app:covariance} we show that 
\begin{enumerate}
    \item The theoretical and numerically regularized jackknife covariance matrices yield consistent parameter constraints.
    \item The posterior is not sensitive to the choice of fiducial parameters used to generate the theoretical covariance matrix.  Consequently, holding the covariance matrix fixed in our analysis is well justified.
\end{enumerate}

For satellites, some of the assumptions in the derivation of the theoretical covariance of the entral CLF data break.  Since we have shown that the jackknife resampling method yields parameter constraints consistent with our theory covariance matrix, we rely on the jackknife covariance matrix for the analysis of the satellite CLF.

Finally, we emphasize that while we consider the full covariance matrix of the luminosity function in each redshift and richness bin, we don't consider the cross covariance matrix between redshift and richness bins. Since our results are measured in wide redshift and richness bins, we expect the covariance matrix of the luminosity function between redshift and richness bins to be relatively small compared to the covariance matrix of luminosity function in the same redshift and richness bins. Thus, we set those off-diagonal terms to be zero and leave further treatments of the full covariance matrix to future studies.

\section{Modeling the Conditional Luminosity Function}\label{sec:modeling}

As in numerous previous studies (e.g., \citealt{YMB2005}), we divide the CLF into two parts.  Given a mass of the cluster, the central galaxy distribution $\Phi_c$ is assumed to follow a log-normal distribution. We assume the satellite CLF $\Phi_s$ is described by a modified Schechter function with characteristic luminosity $L^*$, normalization $\phi^*$, faint-end slope $\alpha$, and bright-end slope $\beta$. All together, our CLF model takes the form
\begin{align}
	\Phi(L|M) &= \Phi_c(L|M) + \Phi_s(L|M) ,\\
	\label{eq-cen-gen} \Phi_c(L|M) &= \frac{1}{\sqrt{2 \pi\sigma_{\rm{ln} L}^2}} \exp{\left(-\frac{(\ln L-\ln L_c)^2}{2 \sigma_{\ln L}^2}\right)}, \\
	\label{eq-sat-gen} \Phi_s(L|M) &= \frac{\phi^*(M)}{L^*\ln10} \left(\frac{L}{L^*}\right)^{\alpha}\exp{\left(-\left(\frac{L}{L^*}\right)^{\beta}\right)},
\end{align}
where $L^*$ and $L_c$ are functions of halo mass. 

Note that $\Phi_c(L|M)$ is unit normalized.  This reflects the fact that the survey depth of SDSS is sufficient to resolve the central galaxy of every cluster in our sample. 

\subsection{Richness--mass relation}
 \label{sec-mass-relation}
 
Halo mass is not an observable. To tie the measured luminosity function to the mass of host halos, we adopt the richness--mass relation measured for the same cluster sample by \cite{SDSS_cluster_cosmology}.  Here, we briefly summarize this relation, and discuss how to incorporate it into the model of the conditional luminosity function in the next two subsections.

The richness--mass relation is modeled as the convolution of the two probability distributions:

\begin{equation}
\label{eq:richness-mass}
P(\lambda_{\rm{obs}}|M, z) = \int d\lambda_{\rm{true}} P(\lambda_{\rm{obs}}|\lambda_{\rm{true}}, z) P(\lambda_{\rm{true}}|M) d\lambda_{\rm{true}},
\end{equation}
where the probability  $P(\lambda_{\rm{obs}}|\lambda_{\rm{true}}, z)$ models observational noise and projection effects. The observational noise is measured by injecting $10,000$ synthetic clusters into the SDSS data set of known richness and measuring their recovered richness \citep{Projection_effect}.  In contrast, projection effects depend on the large-scale structure of the Universe and therefore cannot be properly quantified by randomly-injected clusters. Instead, \cite{SDSS_cluster_cosmology} calibrate this effect using N-body simulations populated with galaxies using a simple mass-richness relation. For details of this calibration, we refer the reader to \cite{Projection_effect}.

The second term in the integral accounts for the intrinsic relation between cluster richness and cluster mass. Different parameterizations have been tested by \cite{SDSS_cluster_cosmology}; none had a significant effect on the cosmological constraints. Here, we adopt the model in which $P(\lambda_{\rm{true}}|M)$ is a log-normal distribution with 
\begin{eqnarray}
\label{eq:richness_mass}
&\langle\lambda_{\rm{true}}|M\rangle &= \lambda_0(M/M^*)^{\alpha} 
\nonumber \\ 
&\sigma^2 &= \sigma_{\rm{intr}}^2+(\langle\lambda_{\rm{true}}|M\rangle-1)/\langle\lambda_{\rm{true}}|M\rangle^2,  
\end{eqnarray}
where the pivot point $M^*$ is set to $10^{14.344} h^{-1}M_\odot$.

With this richness--mass relation in hand, we now describe our model of conditional luminosity functions. 

\subsubsection{Central Model}\label{subsubsec-cen}
We first write down the most general model for the central galaxy luminosity function:
\begin{align}
	&\Phi_c(L|\lambda) = \frac{n(L, \lambda)}{n(\lambda)}, \nonumber \\ &n(L, \lambda) = \int_{\ln\lambda_{\rm{min}}}^{\ln\lambda_{\rm{max}}} \int n(M)P(\lambda, L|M)dMd\ln\lambda, \nonumber 
	\\ &n(\lambda) = \int_{\ln\lambda_{\rm{min}}}^{\ln\lambda_{\rm{max}}}\int_{\ln L_{\rm{min}}}^{\ln L_{\rm{max}}} \int n(M) P(\lambda, L|M) dM  d\ln L d\ln\lambda,
\end{align}
where $P(\lambda, L |M)$ is the joint probability of richness and central galaxy luminosity. $n(M)$ is the Tinker halo mass function \citep{Tinker2008} calculated using the HMFcalc tool \citep{MPR2013} at the mean redshift of each redshift bin.

$P(\lambda, L|M)$ can be further decomposed into $P(L|\lambda,M)P(\lambda|M)$, where $P(\lambda|M)$ is the same as equation \ref{eq:richness-mass} evaluated at the mean redshift of each individual redshift bin.

We model $P(L|\lambda,M)$ as 
\begin{eqnarray}
    \label{eq:chto_central}
    P(L|\lambda,M) &=& \frac{1}{\sqrt{2\pi\sigma_{\ln L}^2}}exp(\frac{-(\ln L-\langle \ln L_c| M,\lambda \rangle)^2}{2\sigma_{\ln L}^2}), \nonumber\\
    \langle \ln L_c| M,\lambda \rangle &=& \langle \ln L_{c} | M\rangle+ d_{\rm{eff}}(\ln\lambda-\langle \ln\lambda\rangle(M)),
\end{eqnarray}
where $\langle \ln L_{c} |M \rangle$ is the mean log central galaxy luminosity of halos with mass $M$,  $\sigma_{\ln L}$ is the scatter of log central galaxy luminosity at a fixed halo mass, and $\langle \ln\lambda\rangle(M)$ is the mean log richness of halos with mass $M$, which can be calculated by integrating $\ln\lambda$ over $P(\lambda|M)$ given by equation \ref{eq:richness-mass}.  $d_{\rm{eff}}$ in equation \ref{eq:chto_central} describes the possible correlation between central galaxy luminosity and richness at a fixed halo mass. A positive $d_{\rm{eff}}$ indicates that the luminosity of a central galaxy would be larger than the mean at a fixed halo mass if the richness of the cluster is also above the mean. 

It may seem odd to introduce $d_{\rm{eff}}$, instead of using the correlation coefficient itself; i.e. one may be tempted to replace $d_{\rm{eff}}$ by $r\sigma_{\ln L_c}/\sigma_{\ln \lambda}$. We have opted not to do so because the scatter relation $d_{\rm{eff}}=r\sigma_{\ln L_c}/\sigma_{\ln \lambda}$ is specific to a log-normal model assuming mass-independent scatter. The $d_{\rm{eff}}$ parameterization retains the linear shifts in the expectation values expected from sensitivity of $L_c$ to richness within a more general setting.  In particular, note that $d_{\rm{eff}}=0$ implies that $L_c$ is independent of richness at fixed halo mass.

Finally, following other CLF studies, we further relate the mean log luminosity of red central galaxies $\langle \ln L_{c}|M \rangle$ to the mass and the redshift of their host halos using the a power-law relation, 
\be\label{eq-m-lcen}
	\langle \ln L_{c}|M \rangle = \ln L_{c0} + A_L \ln\frac{M}{M_{\rm{piv}}} +B_L\ln(1+z)
\ee
Here, M is the cluster halo mass, and $M_{\rm{piv}}$ is the pivot mass set to  $1.57\times 10^{14} h^{-1}M_\odot$ through out this paper. $A_L$ and $\ln L_{c0}$ are two parameters that define the power-law relation between the red central galaxy luminosity and the host halo mass. $B_L$ describes the redshift evolution.

We use Markov Chain Monte Carlo (using the {\sc emcee} package provided by \citealt{For2012}) to fit the five parameters in central CLF model, specifically, $\sigma_{\log L}, d_{\rm{eff}}, \log L_{c0}, A_L, B_L$. We run chains using flat priors on each of the parameters, with the condition that $\sigma_{\log L}$ is positive. The model is fit to our measured CLF, which spans four redshift and four richness bins.  The fit parameters are summarized in Table \ref{tab-clf-cen} and are shown in Figure \ref{fig-cenparam}.  As a consistency check, we also perform a fit to each individual redshift bins. The result is shown in Figure \ref{fig-pev-cen} and summarized in Table \ref{tab-clf-cen_ind}. We found no evidence that the simultaneous fits are driven by a single anomalous redshift bin.

\subsubsection{Satellite Model}

As before, the satellite CLF can be written as
\begin{equation}
 \Phi_s(L|\lambda)=\frac{n(L,\lambda)}{n(\lambda)},   
\end{equation}
where $n(L,\lambda)$ is the number density of satellite galaxies of luminosity $L$ in clusters of richness $\lambda$.  Recall the satellite luminosity function of a halo of mass $M$ is assumed to take a Schechter form according to equation~\ref{eq-sat-gen}. We must, however, account for the fact that we bin in clusters of richness; in general, $\Phi_s(L|M) \neq \Phi_s(L|M,\lambda)$.  Since richness is the number of satellite galaxies, it is obvious that $\Phi_s$ and $\lambda$ must be correlated: richer clusters must have more satellite galaxies by definition.  To account for this covariance, we assume that richness correlates with the amplitude of the luminosity function, but not with its shape.  That is, we assume that the luminosity function is always a Schechter distribution with the same faint-end and bright-end slopes for all clusters.  Likewise, the characteristic luminosity $L_*$ depends only on mass.  However, the amplitude of the luminostiy function $\phi_*$ does depend on the cluster richness.  In fact, we expect these two to be nearly perfectly correlated.  Let then $\Phi_s(L|\phi_*)$ be the luminosity function with amplitude $\phi_*$, and let $P(\phi_*,\lambda|M)$ be the probability that a cluster has richness $\lambda$ and satellite amplitude $\phi_*$.  We have then
\begin{align}
\label{eq:sat}
	&\Phi_s(L|\lambda)= \frac{n(L,\lambda)}{n(\lambda)} \nonumber \\ 
   &= \frac{\int_{\lambda_{\rm{min}}}^{\lambda_{\rm{max}}} d\lambda \int d\phi^*  \int dM\ n(M) P(\phi^*, \lambda | M) \Phi_s(L|\phi^*)}{\int_{\lambda_{\rm{min}}}^{\lambda_{\rm{max}}} d\lambda \int d\phi^* \int dM\ n(M) P(\phi^*, \lambda | M) },
\end{align}

Again, we can decompose $P(\phi^*, \lambda | M)$ into $P(\phi^* | \lambda, M)P(\lambda|M)$.  In the limit that $\phi_*$ and $\lambda$ are pefectly correlated, we have
\begin{align}
    P(\phi^* | \lambda, M) &= \delta(\ln\phi^*-\langle \ln\phi^*|M\rangle-B(\ln\lambda-\langle \ln \lambda |M\rangle )), \nonumber\\
    B &=\sqrt{1-\frac{1}{\sigma^2_{\ln\lambda}\langle \lambda|M\rangle}}.
\end{align}

The value of the coefficient $B$ in the above equation is determined by setting the correlation coefficient between $\phi_*$ and $\lambda$ to unity.  Specifically,
\begin{align}
\label{eq:sat1}
    r &= \frac{(\ln\phi^*-\langle \ln\phi^*|M\rangle) (\ln\lambda-\langle \ln \lambda|M\rangle  )}{\sigma_{\ln\lambda}\sigma_{\ln\phi^*}}\nonumber \\
    &=\frac{B\sigma_{\ln\lambda}^2}{\sigma_{\ln\lambda}\sigma_{\ln\phi^*}} = 1.
\end{align}
In solving for the coefficient $B$ above, we assume that the scatter in $\ln \phi_*$ is identical to the scatter in $\ln \lambda$ up to Poisson fluctuations, so that
\begin{equation}
    \sigma_{\ln\phi^*}^2 = \sigma_{\ln\lambda}^2-\frac{1}{\langle \lambda|M\rangle}.
\end{equation}

The end result is that the satellite luminosity function takes the form

\begin{equation}
    	\Phi_s(L|\lambda)= \frac{\int_{\lambda_{\rm{min}}}^{\lambda_{\rm{max}}} d\lambda \int dM\ n(M) P(\phi^*, \lambda | M) \Phi_s(L|\phi^*)}{\int_{\lambda_{\rm{min}}}^{\lambda_{\rm{max}}} d\lambda \int \int dM\ n(M) P(\phi^*, \lambda | M) },
\end{equation}
where $\phi_* = e^{\langle ln\phi_*|M\rangle + \sqrt{1-1/(\sigma^2_{\ln\lambda}\langle \lambda|M\rangle)}(\ln \lambda - \langle \ln \lambda|M\rangle)}$
The expectation value of $\phi_*$ at fixed mass is then given by a powerlaw,
\begin{equation}
\label{eq-phi}\langle \ln\phi^*|M\rangle = \ln \phi_0 + A_\phi \ln \frac{M}{M_{piv}} +B_\phi \ln(1+z)     
\end{equation}
while the characteristic luminosity $L^*$ is a deterministic function of mass,
\begin{align}\label{eq-lst}
	\ln L^*(M) &= \ln L_{s0} + A_s \ln\frac{M}{M_{piv}} + B_{Ls} \ln(1+z).
\end{align}
The faint-end and bright-end slopes $\alpha$ and $\beta$ are independent of host halo mass.

\begin{figure*}
\includegraphics[width=0.95\textwidth]{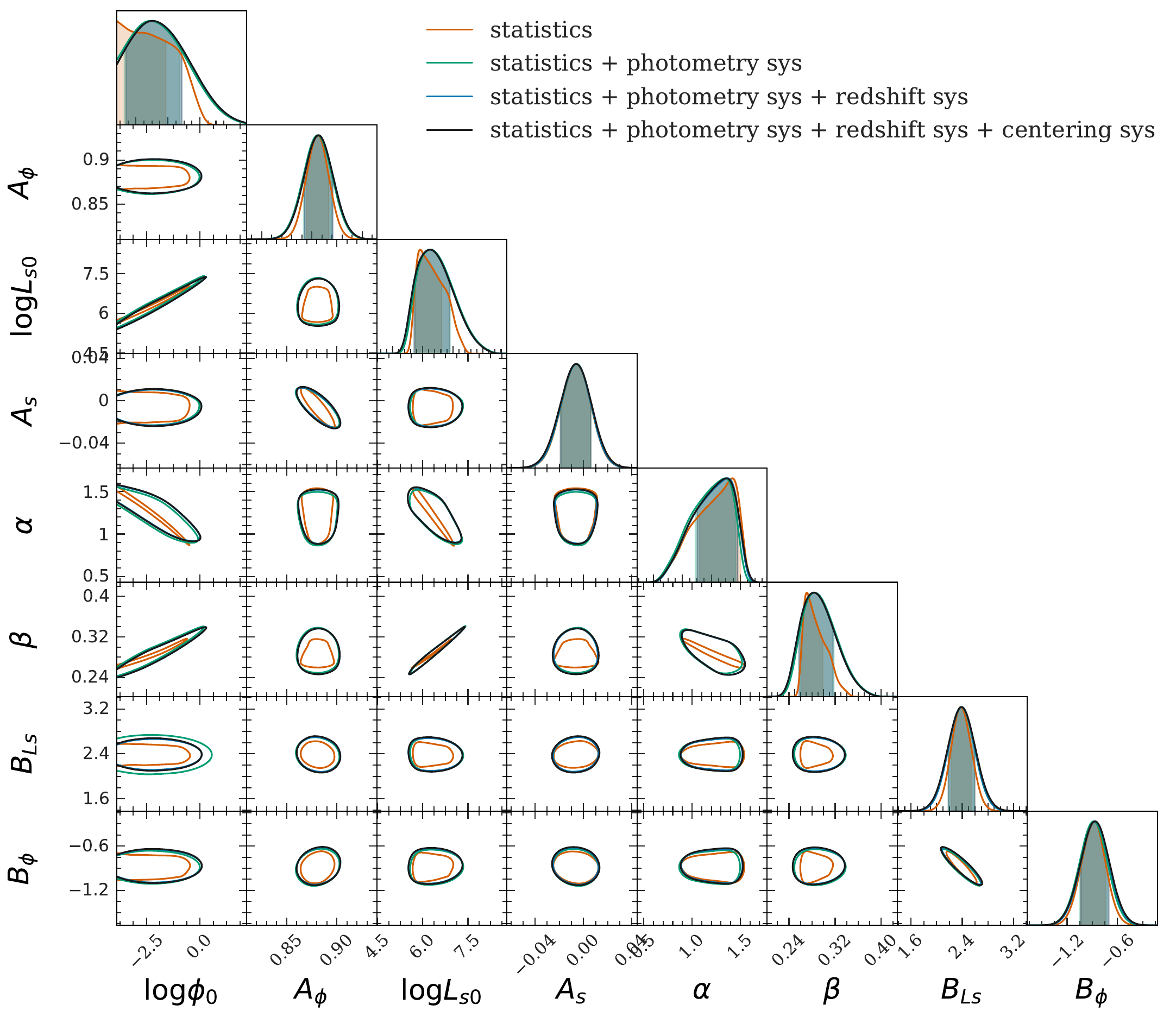}
\caption{68\% contours for the red satellite CLFs. For a list of the best-fit values and the marginalized one--sigma constraints, see Table \ref{tab-clf-sat}. Moving from left to right, the first parameter shown is $\log \phi_0$, the normalization of the red satellite CLF at the pivot mass. $A_\phi$ is the power of the evolution of the normalization $\phi^*$ with host halo mass.  $\ln L_{s0}$ is the characteristic luminosity of the red satellite CLF Schechter function at the pivot mass.  The parameter $A_s$ is the power-law evolution of the characteristic red satellite luminosity with host halo mass.  Note that this value is close to zero.  $\alpha$ is the faint-end slope of the red satellite CLF, while $\beta$ is the bright-end slope of the red satellite CLF. $B_{LS}$ denotes the redshift evolution of the characteristic luminosity $L_{s0}$ and $B_{\phi}$ denotes the redshift evolution of the normalization $\log \phi_0$. Different colors indicate $68\%$ contours including different systematics.}
\label{fig-satparam}
\end{figure*}

Similar to central CLF, we use Markov Chain Monte Carlo ({\sc emcee}, \citealt{For2012}) to fit the eight  parameters of our satellite CLF model, $\log \phi_0$, $A_\phi$, $\log L_{s0}$, $A_s$, $\alpha$, $\beta$, $B_{Ls}$, $B_\phi$. We run the chain using flat priors on each parameter,  except for $\beta$ and $\log \phi_0$. For $\beta$, we require the bright-end slope to be positive and for $\log \phi_0$, we assume a flat prior with range $[-3.9,2.2]$. 
This prior is based on the measurement in \cite{Ber13}, which measured the mean number of galaxies per unit volume and luminosity in the Universe. With this measurement, we conservatively assume that the cluster size is $0.5$-$1.5$ Mpc and that the average galaxy density in a cluster is between $1$ and $1000$ times larger than the galaxy density of the Universe, leading to the $\phi_0$ prior above. It may seem odd that we need a prior on $\phi_0$, but this can be understood as follows.  While $\phi_0$ characterizes the amplitude of the luminosity function, the data only constrains the satellite CLF for $L\geq 0.2L_*$, the luminosity threshold of \redmapper.  This allows for considerable uncertainty in the extrapolation to low luminosities, leading to strong degeneracies between $\phi_0$ and the bright and faint end slopes of the luminosity function (see Figure~\ref{fig-satparam}.)  The prior on $\phi_0$ truncates these degeneracies, preventing us from reaching unphysical conclusions.

We compute the satellite CLF in four evenly spaced redshift bins within $z = 0.1$--$0.3$ and fit the model for all four redshift bins simultaneously. Table \ref{tab-clf-sat} summarizes the fit parameters; the result is shown in
Figure \ref{fig-satparam}.
As a consistency check, we also perform a fit to each individual redshift bins. The result is shown in Figure \ref{fig-pev-sat} and summarized in Table \ref{tab-clf-sat_ind}. We found no evidence of our results being driven by a single anomalous redshift bin.

Appendix \ref{sec-validation} details how we validate our models for both central and satellite galaxies through the use of synthetic mock catalogs.

\begingroup
\renewcommand{\arraystretch}{2.0}
\begin{table*}
    \centering
    \caption{Central Conditional Luminosity Function Parameters}
    \label{tab-clf-cen}
    \begin{tabular}{ccccccc}
        \hline
		Parameters & $\sigma_{\log L}$ & $d_{\rm{eff}}$ & $logL_{0}$ & $A_L$ & $B_L$ & $\chi^2$ (dof)\\ 
        \hline
        Units & $\log L_\odot/h^2$ & - & $\log L_\odot/h^2$ & $\log L_\odot/h^2$ & $\log L_\odot/h^2$ &-\\  
        
        Eq. refs &\ref{eq-cen-gen}& \ref{eq:chto_central} & \ref{eq-m-lcen} & \ref{eq-m-lcen}&\ref{eq-m-lcen} &-\\
		\hline

		 Values & $0.21^{+0.03}_{-0.03}$ & $0.36^{+0.17}_{-0.16}$ & $10.72^{+0.05}_{-0.05}$ & $0.39^{+0.04}_{-0.04}$ & $1.10^{+0.31}_{-0.29}$ &302.8 (252.3)\\

		\hline
    \end{tabular}
\end{table*}
\endgroup
\begingroup
\renewcommand{\arraystretch}{2.0}
\begin{table*}
    \centering
    \caption{Satellite Conditional Luminosity Function Parameters}
    \label{tab-clf-sat}
    \begin{tabular}{cccccccccc}
        \hline
		Parameters & $\log\phi_0$ & $A_{\phi}$ & $\log L_{s0}$ & $A_s$ & $\alpha$ & $\beta$ &$B_{Ls}$&$B_{\phi}$& $\chi^2$ (dof)\\ 
        \hline
        Units & $\log((\log L)^{-1})$ & $\log((\log L)^{-1})$ & $\log L_\odot/h^2$ & $\log L_\odot/h^2$ & - & -&$\log L_\odot/h^2$&$\log L_\odot/h^2$&-\\  
        
        Eq. refs &\ref{eq-sat-gen},\ref{eq-phi}& \ref{eq-phi} & \ref{eq-sat-gen},\ref{eq-lst} & \ref{eq-lst} & \ref{eq-sat-gen}&\ref{eq-sat-gen}&\ref{eq-lst} &\ref{eq-phi}&- \\
		\hline
Values & $-2.22^{+1.39}_{-1.26}$ & $0.88^{+0.01}_{-0.01}$ & $6.24^{+0.65}_{-0.53}$ & $-0.01^{+0.01}_{-0.01}$ & $1.36^{+0.12}_{-0.31}$ & $0.28^{+0.03}_{-0.03}$ & $2.39^{+0.21}_{-0.21}$ & $-0.87^{+0.17}_{-0.17}$ &372.1 (383.6)\\
		\hline
    \end{tabular}
\end{table*}
\endgroup

\section{Systematics}\label{sec-sys}

\subsection{Photometry}
\label{subsec-photometry_sys}

As discussed in section~\ref{subsec-photometry}, we de-bias the SDSS luminosity estimates of bright galaxies by calibrating against \textbf{Pymorph} magnitudes. To derive systematic uncertainty in our CLF parameters associated with the above de-biasing, we repeat our analysis without applying the systematic de-biasing detailed in section~\ref{subsec-photometry}. We adopt half of the shift in the recovered parameters between our analysis with and without a systematic correction as the systematic uncertainty associated with biases SDSS photometry.  Note however that the best-fit values are those reported when applying the correction in section~\ref{subsec-photometry}. We investigate the impact of this correction on the scatter of central galaxy luminosity $\sigma_{\rm{log} L}$ in appendix \ref{app:photo-corr}.

\subsection{Centering performance of \redmapper{} cluster finder}

Although we weight each central galaxy candidate by the probability $P_{cen}$ that the galaxy is the correct cluster center, biases in $P_{cen}$ will necessarily introduce systematic errors into our galaxy luminosity estimates.  \cite{Miscentering} find that about $70\%$ of the \redmapper{} clusters are well centered at the most probable central galaxy reported in \redmapper{}. However, the mean probability of the most probable centrals in \redmapper{} is $86\%$, which indicates that the \redmapper\ centering probability is biased. 
  To quantify the resulting systematic uncertainty, we decrease the largest centering probability of each cluster in \redmapper{} catalog by $16\%$ and increase the second largest centering probability by $16\%$. We remeasure the conditional luminosity function and we refit our model. We then quote half of the difference between the best-fit parameters with and without centering correction as the systematic error. 

We are uncertain of what direction that miscentering shifts our parameters to. For instance, if miscentering happens by randomly centering on satellite galaxies, one would expect the presence of miscentering to lead to an increase in the scatter of the central galaxy luminosity.   However, if miscentering happens preferentially in clusters where the central galaxy is faint, and the incorrectly chosen center is bright, then we would expect miscentering to decrease the apparent scatter in luminosity at fixed halo mass.  Because of the lack of clear directionality, we shift the best-fit value to the middle of the best-fit values from both of our analyses (with and without the $P_{cen}$ corrections).  Note this implies that either limit (no correction, or full correction) is one systematic error away.

\subsection{Other systematics}

The next systematic error we consider is the assumption that the conditional luminosity parameters are independent of cosmological parameters. Future work will fit the mass--richness relation measured by weak lensing analysis as well as conditional luminosity function simultaneously to properly marginalize over cosmological parameters. We quantify this systematic error by computing the difference in the CLF model at the best-fit CLF parameters,  assuming DES Y1 cosmology \citep{DESY1} and Planck cosmology \citep{Planck13}. We find that the difference is at $0.1\%$ and $10\%$ of the statistical error for centrals and satellites respectively; thus, we conclude this systematic is subdominant relative to our total error budget. 

Another possible systematic is the completeness of the survey, particularly as it impacts the faint-end slope of the CLF. However, we don't expect survey completeness is likely to impact our result. Specifically, while we fit a CLF model to the measurement, we restrict the fit region to luminosity above $0.2L^*$, which is equivalent to $\log L = 9.5 \log(L_\odot/h^2)$. At $z=0.3$, this corresponds to $m_i=20.56$, which is somewhat brighter than the magnitude at which the survey becomes incomplete (roughly $m_i\approx 20.8$).  Therefore, we expect the effect of incompleteness on our results to be negligible. 

An obvious possible source of systematic uncertainty is our reliance on photometric cluster redshift estimates.  However, as shown in \cite{Ryk2013}, \redmapper{} redshifts are both highly accurate and precise, with $\sim 0.01$ scatter and an even lower bias.  To quantify this, we shift the mean redshift of the cluster in each bin by $0.01$ and re-do the fitting. We see no difference in the final result and conclude that this systematic is not relevant for our study.. 

\begin{figure}
\includegraphics[width=0.5\textwidth]{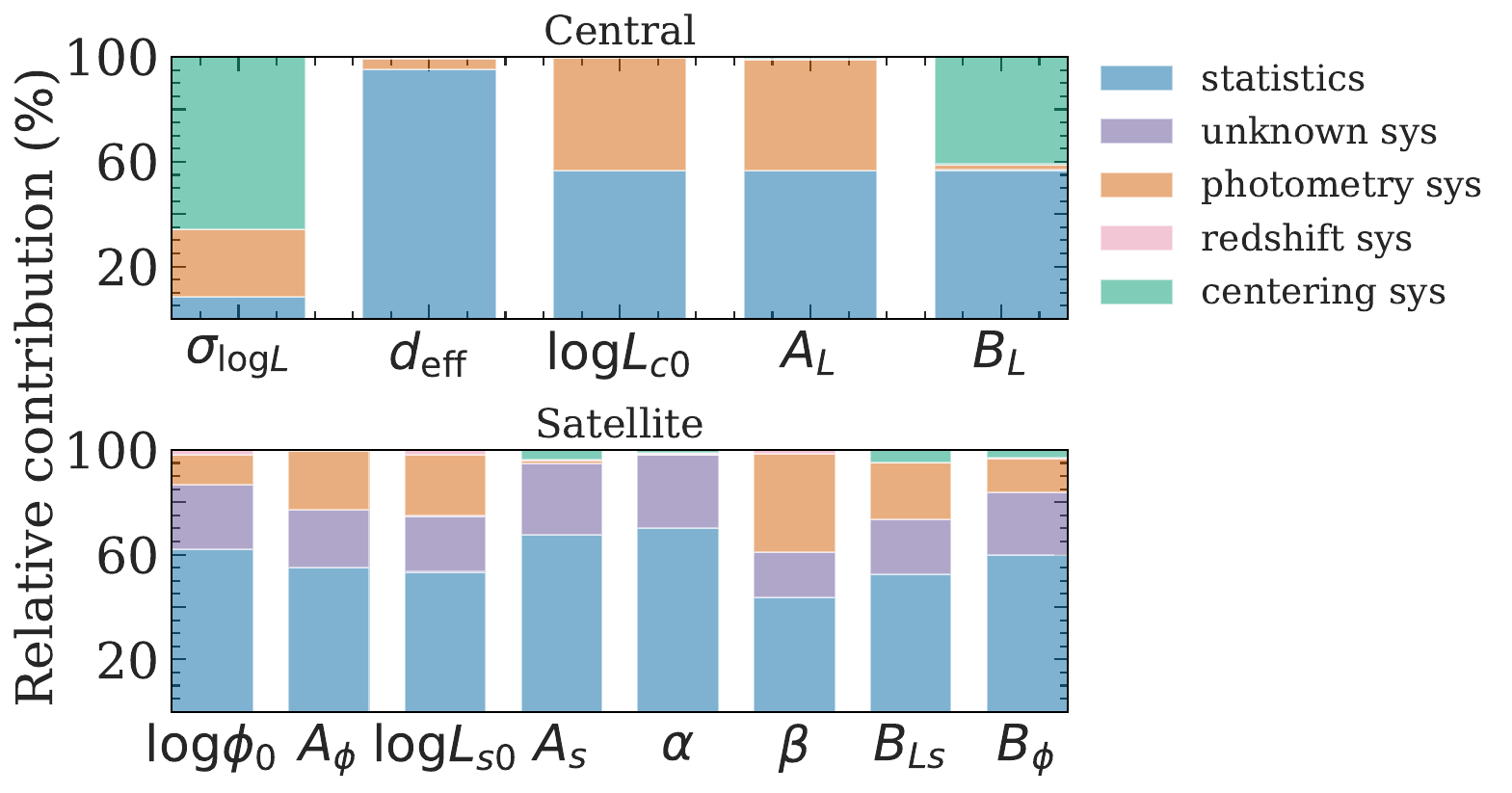}
\caption{Summary of the relative importance of different components in our error budget. Each column corresponds to one parameter of the model. The colors correspond to different components: statistical error (blue), photometry systematic(orange), redshift systematic (pink) and centering systematic (green). For the satellite CLF, the purple region shows the unknown systematics, which we put in by hand as $40\%$ of the statistical error to ensure an acceptable $\chi^2$. } 
\label{fig-sys_relative}
\end{figure}

Combining all systematic errors into our error budget, we assume that all systematics (photometry, centering, and photometric redshift), are mutually independent and independent from the statistical error. With this assumption, we quantify the impact of each type of systematic on the CLF parameters as a Gaussian random variable. The mean is set by the offset of the best-fit CLF parameters due to the relevant systematic and the covariance is set by the product of the parameter correlation function scaled by the systematic error.
 We then draw samples from the multivariate Gaussian distribution and apply them to each step of the MCMC chain before we quantify the marginalized one-sigma error of each parameter. Figure \ref{fig-sys_relative} summarizes the relative importance of different sources of systematics on our error budget for each CLF parameter. 

\begin{figure*}
\includegraphics[width=0.8\textwidth]{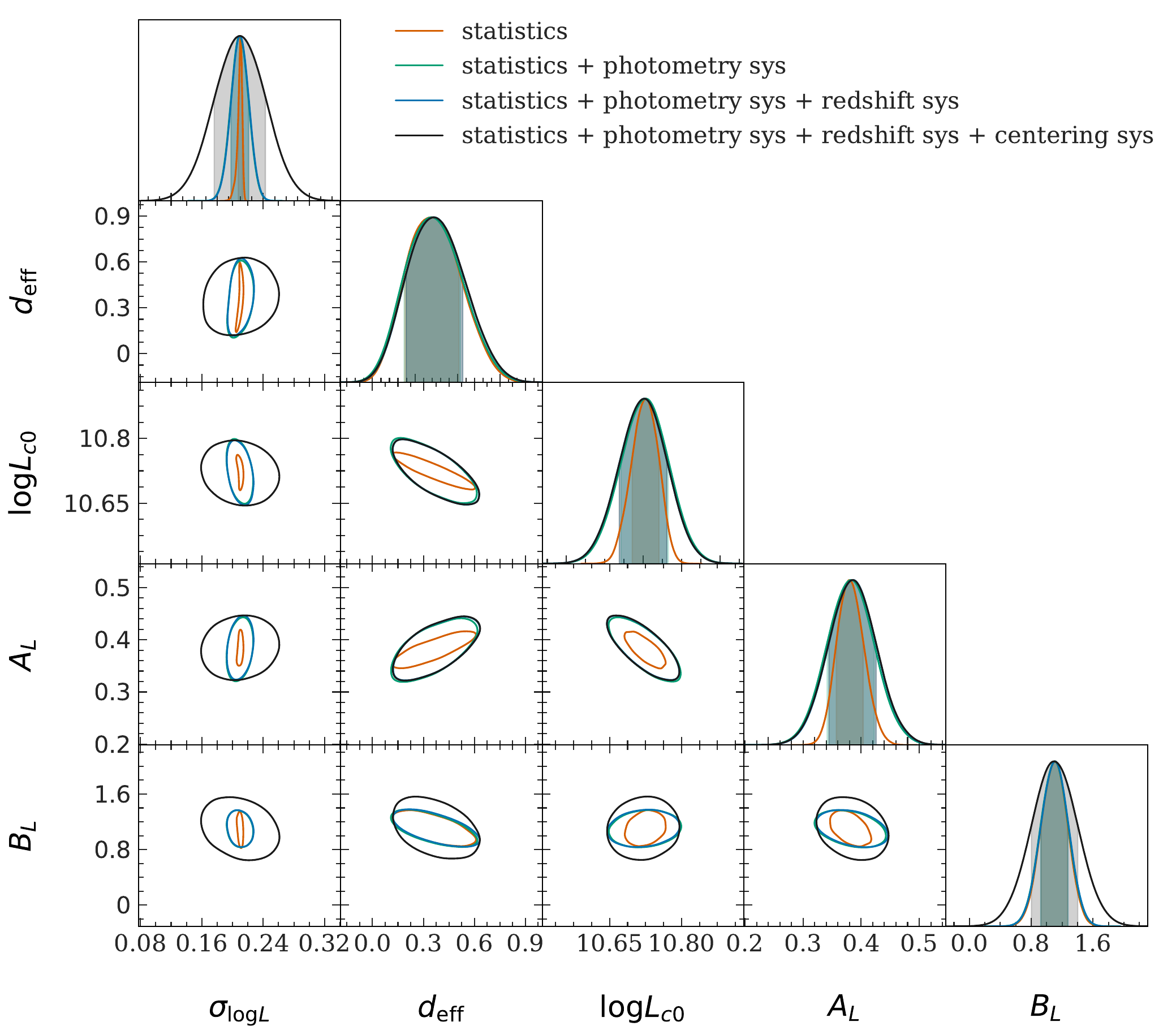}
\caption{68\% contours for the red central CLF parameters. For a list of the best-fit values and the marginalized one--sigma constraints, see Table \ref{tab-clf-cen}.  Moving from left to right, the first parameter shown is $\sigma_{\log L}$, the scatter in dex of the red central galaxy luminosity at fixed host halo mass.  The value of $d_{\rm{eff}}$ describes the possible correlation between the red central galaxy luminosity and the richness $\lambda$ of the cluster at a given host halo mass. Note that $d_{\rm{eff}}$ is correlated with other CLF parameters, and that positive values of $d_{\rm{eff}}$ are preferred. $\log L_{c,0}$ is the typical red central luminosity at the pivot mass, in $L_\odot/h^2$.  $A_L$ gives the power-law relationship between red central luminosity and host halo mass. As expected, this relationship has a significant positive slope. $B_L$ is the redshift evolution of red central galaxies' luminosity. Different colors indicate $68\%$ contours including different sources of systematics.}
\label{fig-cenparam}
\end{figure*}

\section{Results}
\label{sec-ev}

We constrain the conditional luminosity functions (CLF) of red central and satellite galaxies.  Our CLF model for red central galaxies consists of five parameters: $\sigma_{\log L}, d_{\rm{eff}}, \log L_{c0}, A_L, B_L$. The parameter $\sigma_{\log L}$ describes the scatter of red central galaxy luminosity at a given host halo mass, and $d_{\rm{eff}}$ describes the correlation between red central galaxy luminosity and the richness of the host halo. The parameters $\log L_{c0}, A_L$, and $B_L$ describe the power-law relation of mean red central galaxy luminosity and the mass and redshift of their host halos: $\log L_{c0}$ represents the normalization, $A_L$ represents the mass dependence, and $B_L$ represents the redshift dependence. Our CLF model for red satellite galaxies consists of eight parameters, $\log \phi_0$, $A_\phi$, $\log L_{s0}$, $A_s$, $\alpha$, $\beta$, $B_{Ls}$, $B_\phi$. $\log \phi_0$, $A_\phi$, and $B_\phi$ describe the power-law dependence of the normalization of red satellite galaxy luminosity function to the mass and redshift of their host halos. $\log L_{s0}$, $A_s$, and $B_{Ls}$ describe the power-law dependence of the characteristic luminosity of red satellite galaxies to the mass and redshift of their host halos.  When constructing our luminosity functions, we conservatively exclude data below $0.317L^*$, corresponding to $0.2 L^*_{\rm{redm}} + 0.2$ dex, to ensure the satellite samples are complete above this cut. 

The MCMC contours for the model parameters describing red central and satellite galaxies are shown in Figures \ref{fig-cenparam} and \ref{fig-satparam} respectively, with the posteriors summarized in Tables \ref{tab-clf-cen} and \ref{tab-clf-sat}. Our centrals and satellites model yield $\chi^2 = 302.8,$ and $729.3$, with effective degrees of freedom $252.3$ and $383.6$, respectively. We refer the reader to appendix \ref{app:degree_of_freedom} for details on how we determine the degrees of freedom in our fit. 
The fit to the red central galaxies is marginally acceptable, and we leave it as is.  By contrast, the fit to the red satellite galaxy population is not statistically acceptable.  However, the model provides an accurate description of the data, with residuals at the $\approx 12\%$ level.  The bad $\chi^2$ reflects the incredibly small error bars in our measurement. At low luminosities, the error bars in the red satellite luminosity function approach $5\%$.  Since we are content to achieve a description of the data that is accurate at the $\sim 10\%$ level, we simply increase the error bars in the red satellite luminosity function measurements by $40\%$, which leads $\chi^2/dof=1$.  Effectively, we are assuming there are unmodeled effects that explain the modest but statistically significant differences between our model and the data, and we are marginalizing over these effects.

\subsection{Fit for red Central Galaxies}

\subsubsection{The Galaxy Luminosity--Halo Mass Relation of red Central Galaxies}
\label{sec-cen-L_m}

As expected, the red central galaxy luminosity increases moderately with halo mass. Marginalizing over all systematics, we find a slope $A_L=0.39^{+0.04}_{-0.04}$. Since the mass dependence of the red central galaxy luminosity depends on how the luminosity is measured \citep{Zhang2016}, making an apples-to-apples comparison of our results to those in the literature is crucial. We expect the most straight forward comparison we can make is to the measurement in \cite{Kravstov2018}, where they find their measurement of central galaxy magnitude is similar to that in \cite{Meert2015}. \cite{Kravstov2018} find the central galaxy luminosity of a cluster increases with halo mass with a power of $0.4^{+0.1}_{-0.1}$ as determined using 30 X-ray clusters. Assuming the mass-to-light ratio is constant for central galaxies in \redmapper{}--like clusters \citep{Masstolight}, our result is consistent with these results. 

We also measure the dependence of red central galaxy luminosity on redshift. We find $B_L=1.10^{+0.31}_{-0.29}$. Aside from the actual growth of central galaxies, the measurement also contains pseudo-evolution \citep{2013ApJ...766...25D}, the change of halo mass due to the evolution of the mean matter density of the Universe, and passive evolution, the change in galaxy luminosity due to the stellar evolution. We find that the pseudo-evolution contributes $B_L=0.23$ for halos at our pivot mass $M=1.57\times 10^{14} h^{-1}M_{\odot}$, as estimated using the {\sc colossus} package \citep{Colossus} and our best-fit slope $A_L$. To calculate passive evolution, we use the {\sc EZGal} package \citep{EZgal} with a Chabrier initial mass function \citep{Chabrier} and a simple stellar population model (SSP) \citep{SSP1, SSP2} to produce stellar population templates. We assume a formation redshift at $z=2$ with an exponential decaying star formation history (e folding time = $0.1$), and consider a range of metallicity values from $Z=0.05-0.4 \rm{Z_\odot}$ \citep{2000A&ARv..10....1K}. Under these assumptions, passive evolution contributes $B_L=0.57-1.09$ for halos at our pivot mass. After account for both pseudo-evolution and passive evolution, the remaining redshift evolution is characterized by an effective slope $B_L=0.04^{+0.57}_{-0.51}$.  This value is statistically consistent with the determination of \cite{Zhang2016}.

Finally, we measure $\sigma_{\log L}=0.21^{+0.03}_{-0.03}$, with the error bar dominated by the centering systematic (as shown in Figure \ref{fig-sys_relative}). When comparing our result to other values in the literature, it is important to emphasize that this value corresponds to scatter in central galaxy luminosity given mass \it and \rm richness, not just mass.  Under the assumption that the richness--mass relation follows a log-normal distribution with scatter of $0.3$ dex, we can derive the scatter of central galaxy luminosity at a fixed halo mass $\sigma_{\log L|M} = 0.23^{+0.05}_{-0.04}$ \footnote{$\sigma_{\log L}=0.21^{+0.03}_{-0.03}$ if we assume no scatter in richness--mass relation.}.  Although this estimate is statistically consistent with that presented in other studies \citep{Kravstov2018,YMB2009,Cacciato2013}, the scatter we obtained is one to two sigma higher than measurements based on clustering analyses \citep{Cacciato2013, Reddick2013}. This is possibly because the aforementioned studies are based on the Blanton luminosity function \citep{Bla2005b}, which is demonstrated to be too steep at the bright end due to the issue of background subtraction \citep{Ber13}. The underestimated number of galaxies with high luminosity would result in a smaller inferred scatter in those studies and is likely the reason that our analysis infers a larger scatter than studies based on the Blanton luminosity function.  

\subsubsection{Correlation between richness and central galaxy luminosity}

Our model infers a positive correlation between the richness of galaxy clusters and the luminosity of the red central galaxy when holding the halo mass fixed.  The measured effective correlation parameter $d_{\rm{eff}}$ is $0.35^{+0.18}_{-0.16}$. The positive value is hard to explain by observational systematics: one would expect cluster finders would underestimate the number of satellites when the central galaxy is too bright, which would result in a negative correlation. The positive correlation also has the opposite sign as the prediction of halo properties based on concentration dependence. For example, \cite{mao2015} showed that halos with high concentration have fewer satellites, and \cite{Lehmann2017} showed that halos with high concentration should host brighter galaxies to explain the galaxy clustering and satellite fraction measurements. Combining these two, we would have expected the effects of halo concentration would 
lead to a negative correlation between central galaxy luminosity and richness of host halos.  

One possible explanation for this positive correlation is that the projection effect is correlated with the formation history of the halos. Early--forming halos tend to live in denser regions, thus having a larger projection effect that boost the richness $\lambda$ of clusters. Also, in the current paradigm of galaxy formation and evolution, early forming halos tend to have brighter centrals \citep{Lin2013, Zhang2016}. These two effects would manifest as a positive correlation between central galaxies' luminosity and the richness of host halos at a fixed halo mass. This interpretation raises the possibility of enabling us to suppress the impact of projection effects in photometric richness estimation by exploiting the central galaxy luminosity.  We leave pursuing this intriguing possibility to future work.%

\subsection{Fit for red Satellite Galaxies}

We find that the characteristic luminosity $L_*$ describing the red satellite CLF depends only weakly on host halo mass, consistent with the findings of \cite{Hansen09}.  However, we find a strong redshift dependence of the red satellite CLF. Our measured power-law redshift dependence parameter $B_{Ls}$ is $2.39^{+0.21}_{-0.21}$. Similar to the redshift scaling of central galaxies, this value also contains the contribution of psudo-evolution and passive evolution. The combination of these two effects would contribute $B_{Ls}=0.56$--$1.08$. Therefore, our measurement points to red satellites getting dimmer by 25\% to 35\% between $z=0.3$ and $z=0.1$, corresponding to $\sim 2.1$Gyrs of evolution. The dimming of red satellite galaxies can be interpreted as arising from tidal stripping of red satellite galaxies as they fall into clusters.  In related work, \cite{Tidalstriping} used subhalo abundance matching to assign stellar masses to subhalos at infall.  They then compared the resulting galaxy distribution to the conditional stellar mass function at redshift zero to infer that galaxies lose $20-25\%$ of their stellar mass over $\sim 1.3$Gyrs.  The two inferred mass loss rates are comparable.  

We also find the bright-end slope of red satellite luminosity function deviates from a Schechter function. The measured bright-end slope $\beta$ is $0.28^{+0.03}_{-0.02}$, which is consistent with the findings in \cite{Ber13}.

As shown in Figure \ref{fig-dr8clf}, we notice that our red satellite CLF model does not describe the luminosity function well below $\rm{log} L = 9.5 h^{-1}M_{\odot}$. However, since this luminosity range is below our luminosity cut, we can not distinguish between the possibility that it is due to the failure of our model, and the possibility that it is due to incompleteness of the measurement at this luminosity range. We leave further investigation to future.

\section{Discussion}\label{sec-discuss}

We divide our discussion into two sections. We first discuss the inferred mean luminosity halo mass relation and then discuss the relationship between centrals and satellites.

\subsection{Mean luminosity host halo mass relation}
\label{sec:LMrelation}

\begin{figure}
\includegraphics[width=0.45\textwidth]{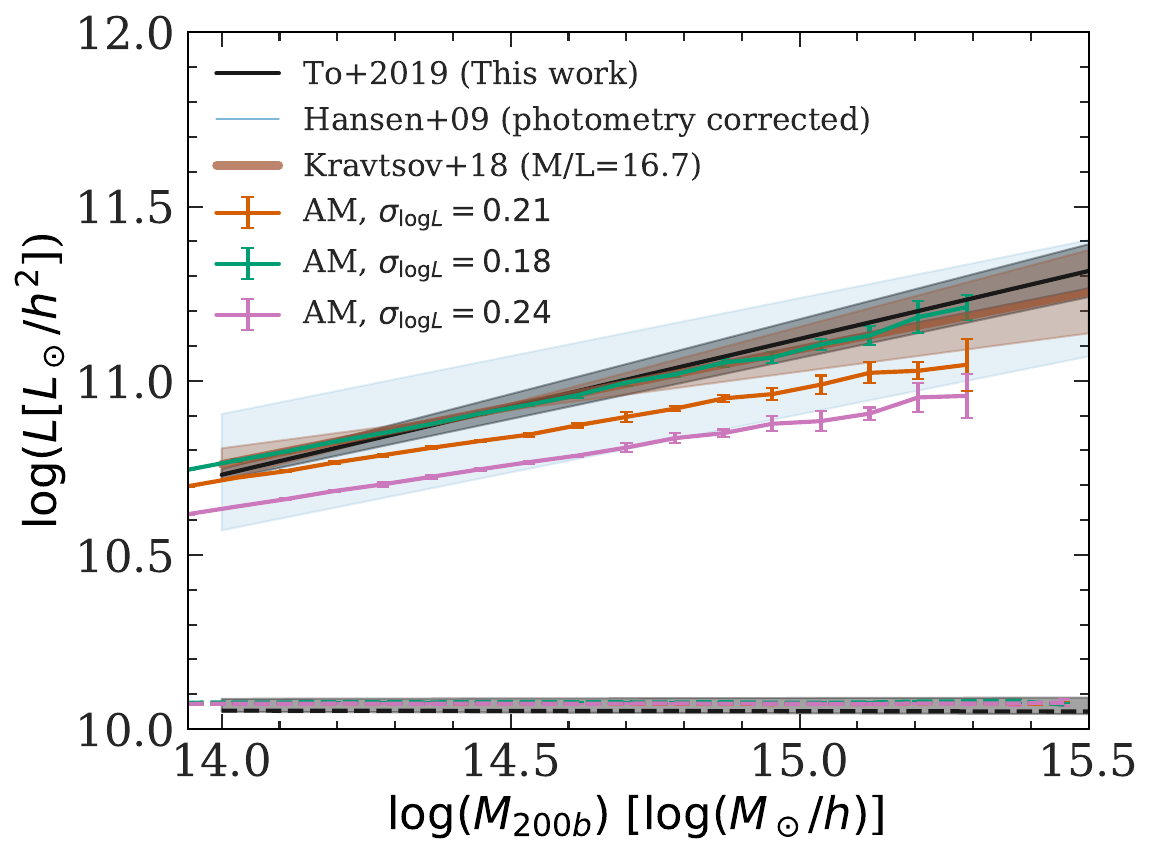}
\caption{Mean luminosity as a function of host halo mass. Solid lines correspond to red central galaxies while the dashed lines correspond to red satellites. While calculating mean luminosity of satellites, we adopt an additional luminosity cut at $\log L = 9.8 L_\odot/h^2$ to avoid the $0.317L^*$ selection while counting satellites. The black line corresponds to the best-fit value inferred from our model with the grey region denoting the one-sigma error. As a comparison, we overplot \cite{Hansen09}'s best-fit value (blue line) with the one-sigma error shown as the blue shaded region. As mentioned in Section \ref{sec-cen-L_m}, different photometry definitions result in different luminosity--mass relations. To make an apples-to-apples comparison, we apply the correction described in Section \ref{subsec-photometry} to \cite{Hansen09}'s best-fit value. We also overplot \cite{Kravstov2018}'s result as brown line with shaded region denoting one-sigma error. Since \cite{Kravstov2018}'s measurement uses stellar mass and ours uses luminpsity, we assume a constant mass-to-light ratio to overplot their result. The mass-to-light ratio is chosen so that \cite{Kravstov2018}'s measurement matches our measurement at $M_{\rm{halo}} = 10^{14.5} M_{\odot}/h$. Finally, we overplot the prediction of subhalo abundance matching with three  values of  scatter corresponding to the best-fit (red), upper-one-sigma (green), and lower-one-sigma (magenta) values of our model. We find our result is in general consistent with the subhalo abundance matching prediction.}
\label{fig-lcen-m}
\end{figure}

One of the key results of this paper is an accurate measurement of the relation between mean galaxy luminosity and halo mass. To compute the mean galaxy luminosity, we integrate the conditional luminosity function in this work and in the literature from $\rm{log} L=9.8 h^{-1}M_\odot$ to $\rm{log} L=14 h^{-1}M_\odot$. We compare our result to closely related work \citep{Hansen09} that used a different cluster catalog. To make an apples-to-apples comparison, we apply the photometry correction described in Section \ref{subsec-photometry} to their data, and shift the pivot mass in \cite{Hansen09} by $18\%$ upward to account for the mass bias described in \cite{Rozo09}. As shown in Figure \ref{fig-lcen-m}, we find that our result is consistent with the measurement of \cite{Hansen09} after applying a correction on photometry with the method described in Section \ref{subsec-photometry}.  However, our results properly marginalize over the possible correlation between cluster observables and account for a variety of systematic effects that have not been previously addressed in the literature.

We also compare our results to predictions from subhalo abundance matching (SHAM). We produce SHAM catalogs using the Rockstar \citep{Rockstar} halo catalog of the Multidark Plank \citep{MDPL} 1$h^{-1}Gpc^3$ simulation box. 
We adopt the parameterization of \cite{Lehmann2017}, matching galaxy luminosity to $v_\alpha = v_{vir}(v_{max}/v_{vir})^\alpha$, with $\alpha = 0.57$, the best-fit value in \cite{Lehmann2017}.   Finally, we abundance match the halo mass function to the \cite{Ber13} luminosity function following a process identical to that of \cite{Reddick2013}.

For the abundance matching model, we assume three different values of scatter between galaxy luminosity and $v_\alpha$: $0.21$ (the best-fit value of the central CLF), $0.18$ (one sigma low), and $0.24$ (one sigma high). As shown in Figure \ref{fig-lcen-m}, the abundance matching result is highly sensitive to the assumed scatter, but is broadly consistent with our measurement. The fact that the subhalo abundance matching results with input from the total galaxy luminosity function match the red central and red satellite CLF data is an interesting, highly non-trivial self-consistency test of both the SHAM framework and our own measurements.

\subsection{Central and Brightest Cluster Galaxies}\label{sec-pbcg}

 Some previous studies have suggested that the brightest cluster galaxy (BCG) may be nothing more than the brightest outlier of the satellite distribution (e.g., \citealt{PS2012}, and references therein). Other work (\citealt{LOM2010, More2012}) indicates that at least in very massive clusters, the BCG is clearly distinct from other galaxies and cannot be defined simply as the brightest galaxy in the population drawing from a single distribution. One natural explanation is that BCGs are central galaxies which follow a luminosity distribution that is distinct from that of the satellites. However, studies have also shown that not all central galaxies are BCGs. For example, \cite{Lange2018} find that $P_{\rm{BNC}}$, the probability that brightest halo galaxy is not a central galaxy is roughly $40\%$. Understanding the origin of this probability is obviously related to our understanding of galaxy formation and evolution, and is a critical component of many cosmology analyses relying on an accurate understanding of the galaxy--halo connection \citep{Lange2018, 2014MNRAS.438.2864L, 2012ApJ...744..159L}.   However, despite many measurements of $P_{\rm{BNC}}$ in the cluster mass regime, measurements don't in general agree with each other. \citet{Ski2011} indicates that, depending on mass, as many as $40\%$ of all BCGs are not located at a cluster's center. Recently, \cite{Lange2018} also found that $\simeq 40\%$ of the BCGs are not the central galaxies of their host halos, and that this fraction is strongly dependent on the host halo mass. However,
 \cite{Hoshino2015} considers the galaxy correlation function by directly counting LRGs in the \redmapper{} catalog. They found a much lower $P_{\rm{BNC}}$ than \cite{Ski2011} and \cite{Lange2018}, with $P_{\rm{BNC}}\simeq 20\%$. 
 One critique of \cite{Hoshino2015} is that they are measuring $P_{\rm{BNC}}$ in richness, not mass, and assuming different correlation between observables could lead to a bias on the constraint of $P_{\rm{BNC}}$. Since our CLF model inferred the luminosity--halo-mass relation by marginalizing possible observable correlations, we can predict the appropriate value for $P_{\rm{BNC}}$ given our data.

\begin{figure}
\includegraphics[width=0.45\textwidth]{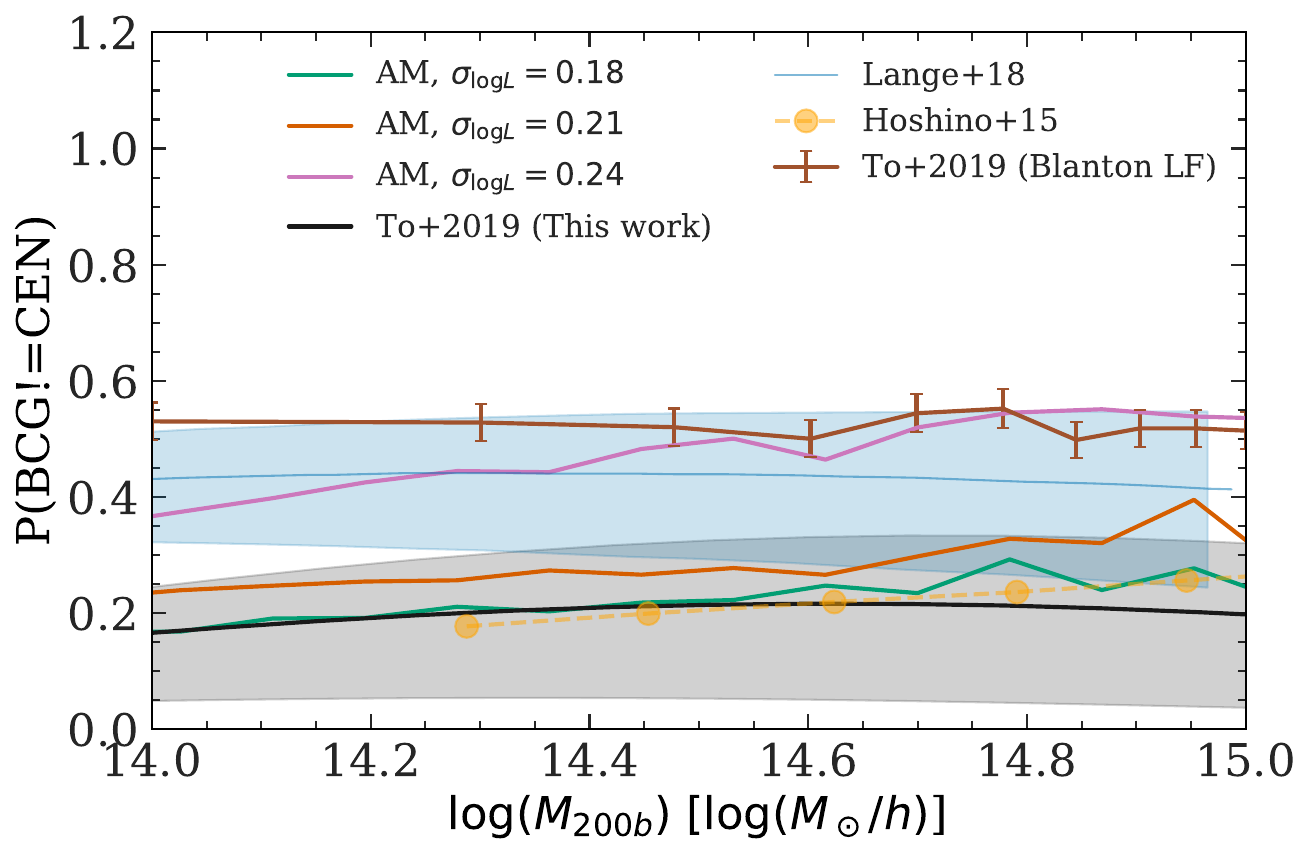}
\caption{Comparison of the probability that the brightest cluster galaxy is not the central galaxy, as a function of host halo mass. The black line corresponds to the predictions from our CLF fits assuming $z=0.2$. The blue line are measurement from \cite{Lange2018}, and the orange dot dashed line are measurement from \cite{Hoshino2015}. We also overplot the prediction of subhalo abundance matching with three values of the scatter corresponding to the best-fit (red), upper-one-sigma (magenta), and lower-one-sigma (green) values of our model. We find that our result is in general consistent with the subhalo abundance matching prediction, but that it is somewhat lower than that found by \cite{Lange2018}. We postulate that the difference of $P_{\rm{BNC}}$ might come from the difference in photometry. To demonstrate this point, we modify the central galaxies' luminosity so that it is consistent to the SDSS DR7 photometry (see Section \ref{sec-pbcg} for details). The corresponding $P_{\rm{BNC}}$ is shown as the brown line, with the error bar representing one-sigma uncertainty.}
\label{fig-pbcg-sham}
\end{figure}
We first define a lower limit of $L_{min}$ for our integrations, where $L_{\rm{min}} \ll L_*$ and $L_{\rm{min}} \ll L_0$, and the specific value of $L_{\rm{min}}$ will not have any significant impact on our results.  Thus, we can determine the expected number of red satellites brighter than $L_{\rm{min}}$:

\begin{align}
	\langle N_s \rangle &= \int_{L_{\rm{min}}}^{\infty} dL \Phi_s(L).
\end{align}

With this as our normalization, we can express the probability distribution for a single red satellite galaxy brighter than $L_{\rm{min}}$ drawn from this Schechter function as:

\be
P_s(L_s) = \frac{\phi^*}{\langle N_s \rangle L^* ln10} \left(\frac{L_s}{L^*}\right)^{\alpha} \exp{\left(-(\frac{L_s}{L^*})^{\beta}\right)}.
\ee

It follows that the probability that this single satellite galaxy is brighter than the central galaxy may be given by:

\begin{align}
	P(L_s > L_c) &= \int_{\log L_c}^{\infty} P_s(L) dL. 
\end{align}

Next, we must consider the case of a cluster, which has $N$ red satellite galaxies.  The probability that at least one of these satellites is brighter than the central galaxy is given by:

\begin{align}
	P( \ge1~L_s > L_c|N) &= 1-\left(1- P(L_s > L_c)\right)^N,
\end{align}

which is one minus the probability that all red satellites are dimmer than the central.

Finally, we must also take account for the fact that our earlier fits implied a correlation between the central luminosity and the cluster richness $\lambda$, which will be proportional to the number of red satellite galaxies in the cluster.
We expect the probability of having a central galaxy in the cluster within the mass range considered here to be very close to 1. Therefore, the number of red satellite in each clusters is $\lambda - 1$. With all the pieces together, our final $P_{BNC}(M)$ reads:

\begin{flalign}
P_{BNC}(M) = \int_{\ln L_{\rm{min}}}^\infty & d\ln L_c  \nonumber \\ &\sum_{\lambda=1}^\infty P( \ge1~L_s > L_c|\lambda-1, M) P(\lambda,L_c|M).
\end{flalign}

Figure \ref{fig-pbcg-sham} shows the $P_{BNC}$ predicted by our CLF model compared to published work. Our value for $P_{BNC}$ is $\approx 10\%-20\%$ lower than that of \cite{Lange2018}.  Nevertheless, due to the large uncertainties in both measurements, the two values are statistically consistent with one another. One of the main differences between our analysis and their work is that we adopt an empirical correction to the central and satellite galaxy's luminosity due to biases in the SDSS magnitudes of bright galaxies. This correction makes central galaxies even brighter (Figure \ref{fig-bernardi_redmapper_M}), and therefore tends to make $P_{BNC}$ smaller. To demonstrate this point, we construct a map from \textbf{PyMorph} luminosity to the SDSS DR7 luminosity by abundance matching Bernardi luminosity function \citep{Ber13} to Blanton luminosity function \citep{Bla2005b}. We then draw $500$ Monte Carlo realization of galaxy clusters for each mass bin according to our CLF model. We then modify the central galaxies' luminosity based on the map we constructed and measure $P_{\rm{BNC}}$. The result is shown as brown line in Figure \ref{fig-pbcg-sham}. Furthermore, \cite{Lange2018} adopted a CLF which assumes that $L^*_{sat}(M)=0.562L_{cen}(M)$. This implies that the ratio of $A_{L}$ and $A_{s}$ should be $1.78$. However, in our best-fit model, this ratio is much larger. The fact that central galaxies are relatively brighter than red satellite galaxies in higher mass halos makes $P_{BNC}$ smaller for high mass halos. 

We also check our results against the SHAM prediction as implemented using the method described in Section \ref{sec:LMrelation}. We find that our measurement is consistent with the abundance matching prediction, another reassuring instance of internal self-consistency.

\section{Summary and Conclusions}\label{sec-summary}
We derive a model for the red central and satellite conditional luminosity function of \redmapper\ clusters, and use it to analyze the SDSS \redmapper\ cluster catalog. The large number of SDSS \redmapper{} clusters and the relatively well-understood richness--mass relation enables a detailed analysis of the conditional luminosity function, which then yields a tight constraint on the galaxy luminosity--halo mass relation. Here we highlight a few of the unique features of this paper compared to the existing literature. First, our model takes into account possible correlations between galaxy luminosities and richness of \redmapper{} clusters at a given halo mass. Second, we consider full bin-to-bin covariance matrices of conditional luminosity functions while deriving the likelihood of this analysis. Third, we incorporate a correction to SDSS DR8 photometry to make it consistent with results from \cite{Ber13}. Forth, our error bar accounts for what we expect are the primary systematics in \redmapper{} cluster samples: photometric biases, centering error, and cluster photometric redshift uncertainty.

Our main results and conclusions can be summarized as follows: 
\begin{enumerate}
	\item The characteristic luminosity $L^*$ of red satellites is very weakly dependent on host halo mass, whereas the central galaxy luminosity increases significantly with host halo mass, with a power-law slope of $~\sim 0.39\pm{0.04}$. This is consistent with the findings in \cite{Hansen09} and \cite{Kravstov2018} but with higher precision.
	
	\item We measure the scatter of central galaxy luminosities at fixed $M_{200b}$, finding $\sigma_{\rm{log}L|M} = 0.23^{+0.05}_{-0.04}$.  This is constrained over the mass range $M_{200b}\sim 10^{14} h^{-1}M_\odot - 10^{15} h^{-1}M_\odot$. 
	
	\item We infer a positive correlation between central galaxy luminosity and the richness of host halos at a fix halo mass. We measure the effective correlation $d_{\rm{eff}}=0.36^{+0.17}_{-0.16}$. This positive correlation increases the mass dependence of the central galaxy luminosity relative to a model in which this correlation is absent.
	
	\item The redshift evolution in the luminosity of central galaxies is consistent with the expectations of pseudo-evolution + passive evolution.  By contrast, red satellite galaxies are dimmed by an amount that is clearly in excess to that predicted by those two effects alone.  We interpret this dimming as evidence of tidal stripping of red satellite galaxies.

	\item The probability $P_{BNC}$ that a cluster's brightest galaxy is not the central galaxy is roughly $20\%$. We note that this inference is sensitive to the photometry of bright galaxies.  
	
	\item We quantify the dominant systematics in this analysis and summarize their relative contribution to our final error budget (Figure \ref{fig-sys_relative}).
\end{enumerate}

In future work, we expect to expand the measurements of these samples. In particular, an examination of the radial distribution of galaxies in clusters (e.g., \citealt{Hansen09,Bud2012,Tal2013}) will help understand the processes surrounding the accretion of satellite galaxies.  This cluster catalog may also be used to investigate the magnitude gap \citep{Tav2011,Hea2012, Dea2013} and how the central and brightest satellite galaxy are related to each other.  The magnitude gap has previously been associated with the assembly history and formation time of the host halo, and may in turn provide access to these properties of \redmapper{} clusters. Moreover, a similar analysis of this paper can be done on \redmapper{} clusters identified in the Dark Energy Survey (DES). With deeper images, such an analysis will provide powerful constraints on the evolution of the galaxy luminosity--halo mass relation, thus shedding light on mechanisms of cluster formation. Finally, the cluster cosmology analysis usually assumes that galaxy clusters correspond to dark matter halos in the simulation. However, it is not hard to believe that the performance of the optical cluster finder depends on the properties of red galaxies in massive halos. That is to say, the amount of \redmapper{} clusters at a given redshift and richness might depend not only on cosmological parameters but also on the parameters of conditional luminosity function. It is then natural to jointly constrain conditional luminosity function parameters as well as cosmological parameters in the cluster abundance analysis. We believe with the recent development of emulator techniques, such an analysis is feasible in the near future. 

\section*{Acknowledgements}

We thank Susmita Adhikari, Albert Chuang, Gregory Green, Daniel Gruen, John Moustakas, and Yuanyuan Zhang for helpful discussions. We thank the anonymous referee for a number of helpful comments that improved the manuscript.

This work was supported in part by the U.S. Department of Energy contract to SLAC no. DE-AC02- 76SF00515, by the National Science Foundation under NSF-AST-1211838, and by Stanford University.  RR was supported by a Stanford Graduate Fellowship. ER was supported by the DOE grant DE-SC0015975, and by the Cottrell Scholar program of the Research Corporation for Science Advancement. Part of this work was performed at Aspen Center for Physics, which is supported by National Science Foundation grant PHY-1607611. Most of the computing for this project was performed on the Sherlock cluster. We would like to thank Stanford University and the Stanford Research Computing Center for providing computational resources and support that contributed to these research results.

This work used data from the SDSS survey; funding for SDSS-III has been provided by the Alfred P. Sloan Foundation, the Participating Institutions, the National Science Foundation, and the U.S. Department of Energy Office of Science. The SDSS-III web site is http://www.sdss3.org/.
SDSS-III is managed by the Astrophysical Research Consortium for the Participating Institutions of the SDSS-III Collaboration including the University of Arizona, the Brazilian Participation Group, Brookhaven National Laboratory, Carnegie Mellon University, University of Florida, the French Participation Group, the German Participation Group, Harvard University, the Instituto de Astrofisica de Canarias, the Michigan State/Notre Dame/JINA Participation Group, Johns Hopkins University, Lawrence Berkeley National Laboratory, Max Planck Institute for Astrophysics, Max Planck Institute for Extraterrestrial Physics, New Mexico State University, New York University, Ohio State University, Pennsylvania State University, University of Portsmouth, Princeton University, the Spanish Participation Group, University of Tokyo, University of Utah, Vanderbilt University, University of Virginia, University of Washington, and Yale University.

\facility{Sloan}
\software{Python, Matplotlib \citep{Hunter:2007}, NumPy \citep{numpy}, GetDist (\url{https://getdist.readthedocs.io/}), Emcee \citep{For2012}, Abundance Matching (\url{https://bitbucket.org/yymao/abundancematching/}), IDL.}

\newpage
\appendix 

\section{Impact of the photometry correction on $\sigma_{\log L}$  }\label{app:photo-corr}

In section \ref{subsec-photometry_sys}, we don't shift the best-fit parameters to account for the systematic due to photometry, since we believe the parameters obtained after the photometry correction are closer to the truth. It is clear that the mean value of the luminosity--halo mass relation is closer to the truth after we apply the photometry correction using the method described in section \ref{subsec-photometry}.  However, whether such a correction also corrects the scatter in the luminosity--halo mass relation is not obvious. Therefore, in this section, we build a toy model to demonstrate that the scatter $\sigma_{\log L}$ in galaxy luminosities with the correction described in Section \ref{subsec-photometry} is closer to the true value of $\sigma_{\log L}$ compared to the one without such a correction. 

First, we generate $7016$ fake central galaxies with absolute magnitude $M_{true}$ drawn from a Gaussian distribution with mean $= -23.5$ and scatter $=0.2$ dex. Then, we mimic the observational effect by adding a random variable $d$ to describe the difference between the observed absolute magnitude $M_{obs}$ and the true absolute magnitude $M_{true}$. Inspired by Figure \ref{fig-bernardi_redmapper_M}, we assume that $d$ is a random variable with mean $=-0.1M_{true}-2.10$ and scatter $0.01$. The scatter is set to the median of the $i$-band one-sigma uncertainty of central galaxies in our sample. The observed magnitude is then defined as $M_{obs} = M_{true}+d$.

Secondly, we fit a linear function to $M_{obs}-M_{true}$ vs $M_{obs}$ relation to mimic the process of the correction described in Section \ref{subsec-photometry}. To be consistent, we randomly select $1516$ galaxies out of $7016$ fake central galaxies to obtain this correction.  We then apply this correction to $M_{obs}$ to obtain  $M_{corrected}$. We measure the scatter of $M_{corrected}$ and $M_{obs}$ to see which is closer to the scatter of true magnitude $0.2$. 

\begin{figure}
\includegraphics[width=0.7\textwidth]{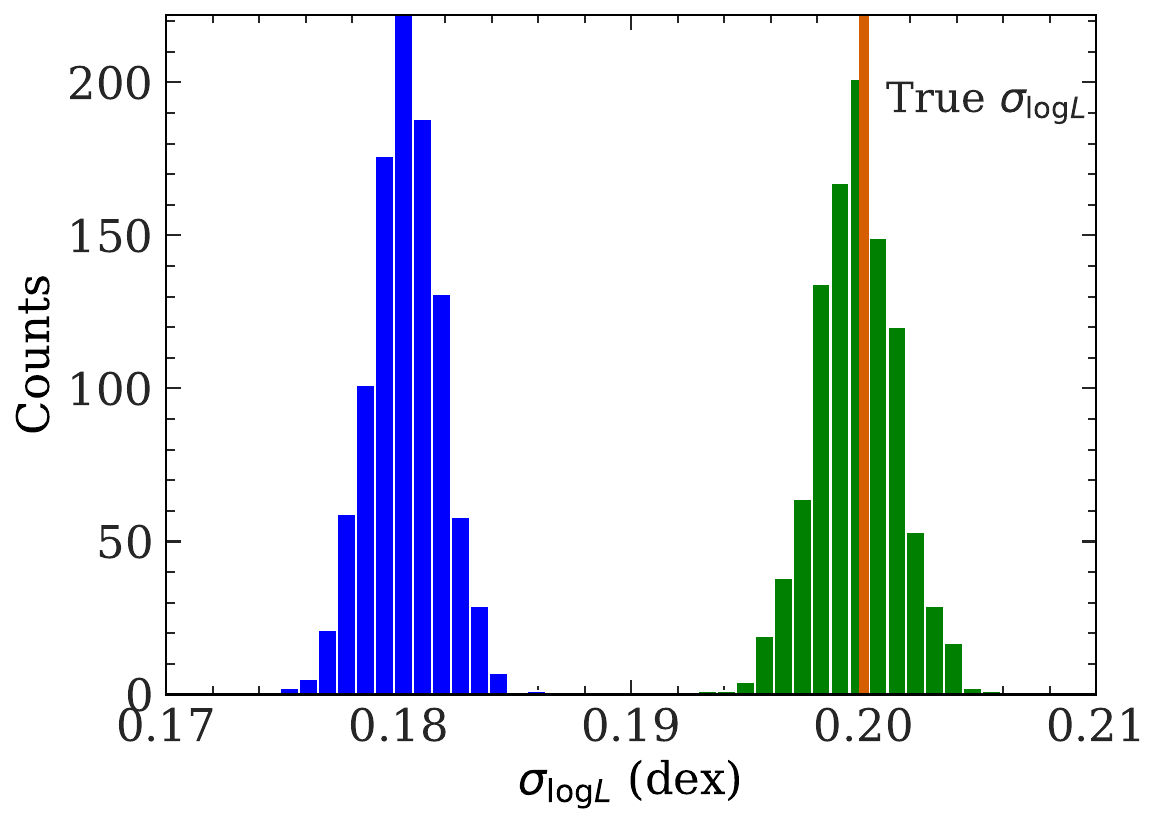}
\caption{Histogram of the scatter of central galaxies luminosity generated by 1000 independent realizations from the procedure described in Appendix \ref{app:photo-corr}.  Green and blue histograms show the scatter of central galaxies luminosity with and without the photometry correction. The orange vertical line indicates the true $\sigma_{\rm{log}L}$ in the simulation described in Appendix \ref{app:photo-corr}.}
\label{fig-Bernardi-sigmaL}
\end{figure}

We repeat the above procedure $1000$ times and show the result in Figure \ref{fig-Bernardi-sigmaL}. As shown in the figure, we demonstrate that the photometry error can leads to a bias on $\sigma_{\log L}$, and our correction can fix this bias. %

\section{Validation of the analysis pipeline using synthetic catalogs}
\label{sec-validation}
We validate our analysis framework by placing CLF constraints on a set of synthetic catalogs, whose CLF parameters are known {\it a priori}. We decide not to use an $N$-body simulation because our model doesn't include clustering of clusters. Thus, a randomly distributed halo catalog is sufficient for this validation. By generating halos from a defined halo mass function, we are able to produce a large set of simulations, thereby making this validation test more statistically significant. 

To construct the synthetic catalogs, we first generate halos from a Poisson draw of the Tinker halo mass function \citep{Tinker2008} and randomly place those halos on a 10405 deg$^2$ sky. We then assign a true richness to each halo using a log-normal distribution, and calculate the observed richness using the $P(\lambda_{obs}|\lambda_{true})$ relation described in Section \ref{sec-mass-relation}. For each halos we populate the central galaxy luminosity using a log-normal model with mean following the power-law relation as described in equation \ref{eq-cen-gen}. We then populate satellite luminosities using equation \ref{eq-sat-gen}. Note that by populating halos this way, we implicitly assume there is no correlation between central galaxies' luminosity and the richness of a cluster at a fixed halo mass. Therefore, this validation test also serves as a null test of the analysis pipeline. 

We generate 100 simulations with the parameters: $\sigma_{\log L}, d_{\rm{eff}}, \log L_{c0}, A_L, A_s, \log L_{s0}, \alpha, \beta = [0.254, 0, 10.722, 0.318, 10.222,-1.084, 0.974]$. The satellite parameters $\log \phi$ and $A_\phi$ are derived from the constraint that the richness of a cluster equals the number of satellites in the cluster plus one. 

Given the synthetic catalogs, we measure the conditional luminosity function following the same procedure described in section \ref{sec-clf}. To avoid statistical noise and to put a stringent test on our analysis pipeline, we use the mean of the measurements on 100 simulations as our data vector and adopt the theory covariance matrix (appendix \ref{app:covariance}) to calculate likelihoods. 

\begin{figure}
\includegraphics[width=0.7\textwidth]{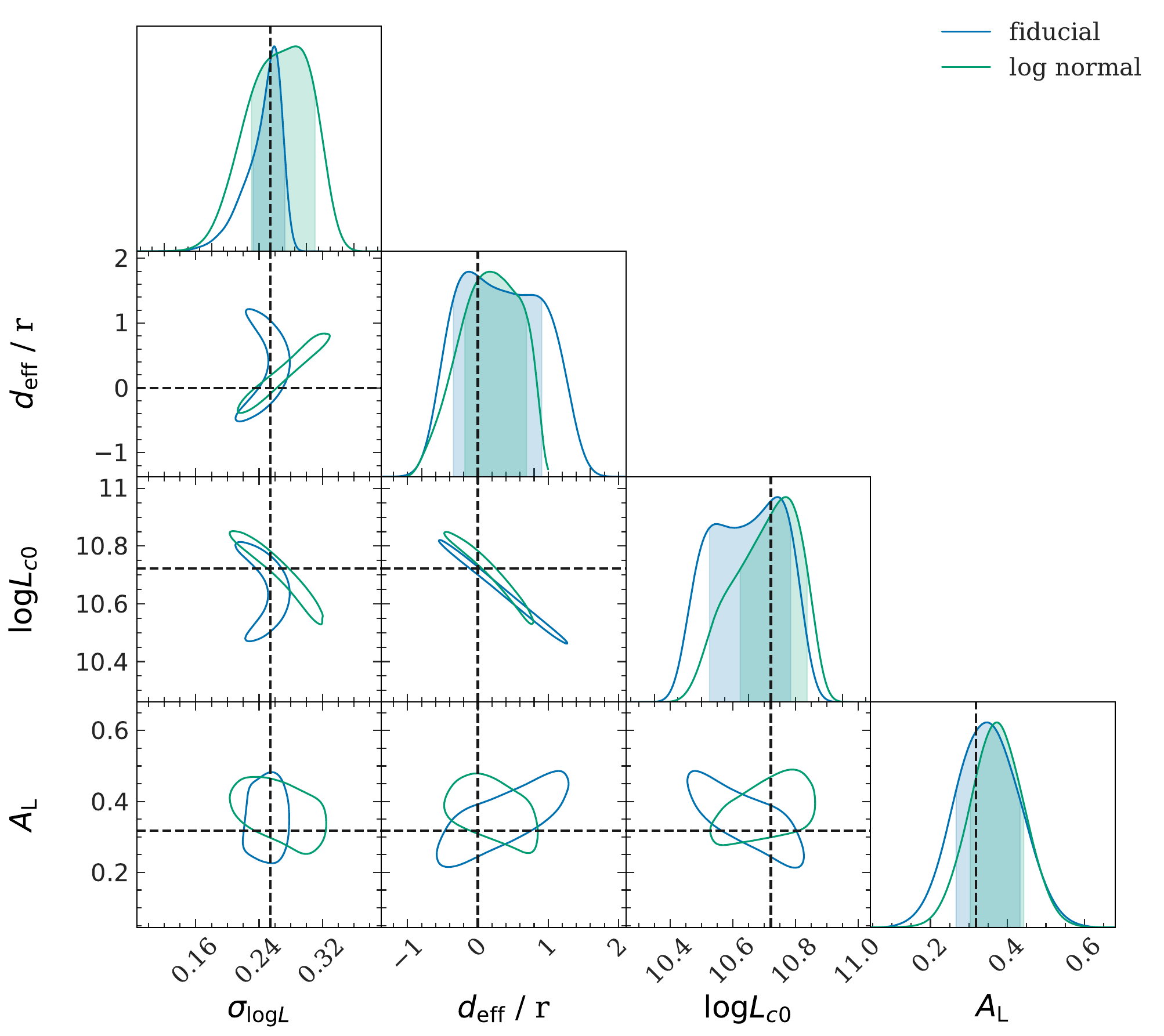}
\caption{$68\%$ parameter constraint (blue) obtained by running our analysis on mock data generated following procedure described in Appendix \ref{sec-validation}. The input parameters of generating these mocks are shown as black dashed line.  As a comparison, we overplot the contours by assuming the $\lambda-M$ relation following a log-normal distribution with constant scatter plus a Poisson term. For the second paramter, we plot $d_{\rm{eff}}$ for our fiducial model, and $r$ for a model assuming $P(\lambda, L|M)$ following a multivariate log-normal distribution.} 
\label{fig-sim-cen}
\end{figure}

\begin{figure}
\includegraphics[width=0.7\textwidth]{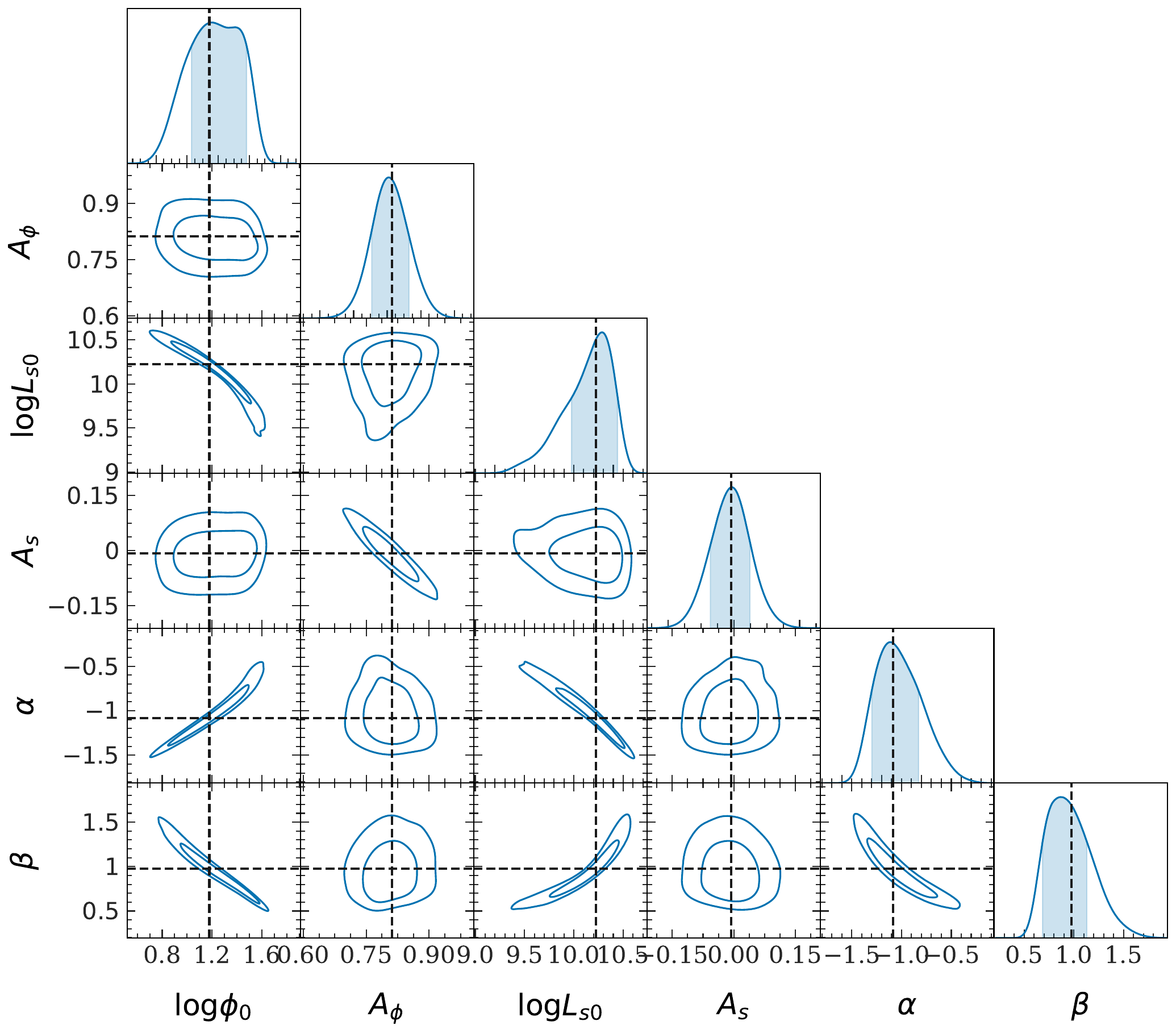}
\caption{Same as Figure \ref{fig-sim-cen}, but for satellites.} 
\label{fig-sim-sat}
\end{figure}

Figures \ref{fig-sim-cen} and \ref{fig-sim-sat} show the constraints we obtained by running our analysis pipelines on the synthetic data for centrals and satellites respectively. We find that our fiducial model successfully recovers our input CLF parameters. 
In Figure \ref{fig-sim-cen}, we also fit our synthetic catalogs with a model assuming that the richness--mass relation follows a log-normal distribution. We first constrain the richnness--mass relation by fitting a log-normal distribution. We then modify the $P(\lambda,L|M)$ term in our model to be a multivariate log-normal distribution with a correlation coefficient $r$. We find that the log-normal model leads to a bias of the posterior, especially on $r, \log L_{c0}$ panel. To better quantify this bias, we approximate the posterior distribution by a multivariate normal distribution and calculate the probability of a random draw having a larger posterior than the posterior evaluated at the input CLF parameters. We find that this probability is $0.211$ for our fiducial model and $0.999$ for the log-normal model, indicating a significant bias on the posterior introduced by the assumption that the richness--mass relation follows a log-normal distribution.  

\section{Model's Degrees of Freedom}
\label{app:degree_of_freedom}
\begin{figure}
\includegraphics[width=0.7\textwidth]{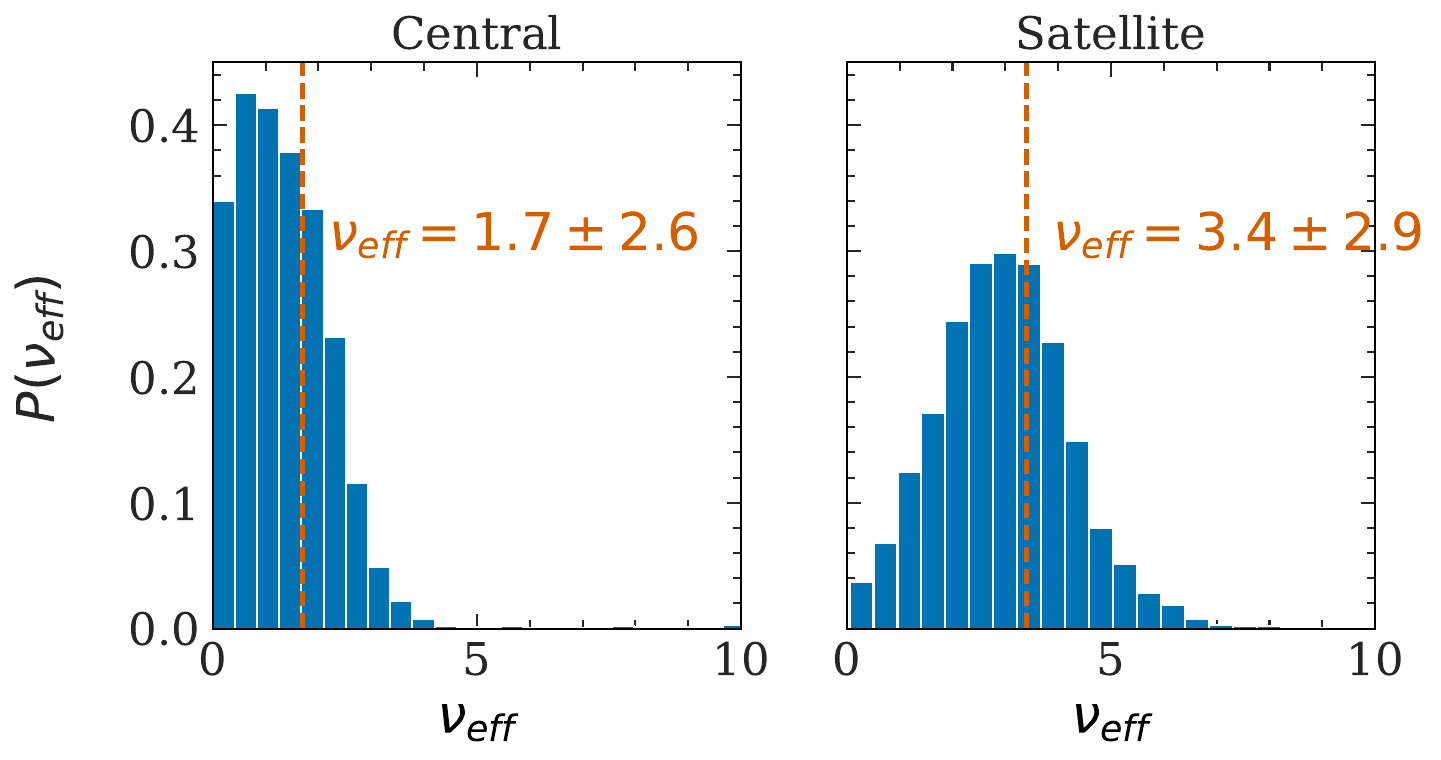}
\caption{Posterior distribution of the number of effective parameters of central (top panel) and satellite (bottom panel) models without redshift evolution parameters ($B_L$, $B_{\phi}$, $B_{Ls}$). The histogram is obtained by fitting $\chi^2$ for 100 mocks generated from the best-fit value.} 
\label{fig-dof_model}
\end{figure}

We calculate the degrees of freedom in the analysis to assess the goodness-of-fit of the model. Normally, the degrees of freedom are defined as the length of the data vector minus the number of parameters in your model. As pointed out in \cite{2010arXiv1012.3754A}, this is no longer true for nonlinear models, as is the case for this paper. To quantify the effective degrees of freedom, we generate $100$ mocks from the best-fit model. For each mock, we apply a mask to the mock data vector and fit the model to obtain the minimum $\chi^2$. Because we fit the same model to different mocks, we expect the number of effective free parameters of the model to stay the same. To constrain the number of effective free parameters, we use the likelihood, 
\begin{equation}
\log \mathcal{L}(\nu_{\rm{eff}}) = \sum_{i} log \scalebox{1.5}{$\chi$}^2 (\chi^2_i, N_i -\nu_{\rm{eff}}), 
\end{equation}
where i runs through 100 mocks. $\scalebox{1.5}{$\chi$}(x, b)$ indicates the probability of $x$ for a chi-squared distribution with degrees of freedom $b$. $\chi^2_i$indicates the least $\chi^2_i$ for mock i, while $N_i$ is the number of datapoints. $\nu_{\rm{eff}}$ denotes the number of effective parameters of the model. 
Figure \ref{fig-dof_model} shows the result for centrals and satellites. In the main text, the effective number of degrees of freedom is used to calculate the degree of freedom to quantify the goodness-of-fit of the model, with the degrees of freedom equal to the number of datapoints subtracted by $\nu_{\rm{eff}}$. 
\section{Covariance matrix}
\label{app:covariance}
\subsection{Theoretical covariance matrix of central CLF}
Here we derive the bin-to-bin covariance matrix for the luminosity function in a single redshift and richness bin. This covariance matrix is adopted in our fiducial analysis.  
To derive the covariance matrix, we follow the formalism of the halo model, in which galaxies are located in dark matter halos. Dark matter halos are then biased tracers of the dark matter density fields, with bias depending on their mass. The dark matter density fields are gaussian realizations of the matter power spectrum which depends on cosmological parameters.  To simplify our analysis, we further make the following assumptions:
\begin{enumerate}
    \item Each galaxy cluster has one and only one central. 
    \item The properties of galaxies only depend on the physical properties of the host galaxy cluster in which they reside. 
    \item The number of galaxy clusters per volume per richness bin follows a Poisson distribution. 
    \item Each galaxy cluster is associated with a unique dark matter halo.
\end{enumerate}

Following equation \ref{eq:measurement}, we then write the central galaxy conditional luminosity function estimator as 
\begin{equation}
\label{eq:estimator}
\hat{\phi}(L_\mu, \lambda) = \frac{\sum_i^{N_{cl}} N^g_i(L_\mu)}{N_{cl}(\lambda)\Delta L_\mu},
\end{equation}
where $N_{cl}$ is the number of galaxy clusters, $N^g_i(L_\mu)$ is the number of galaxy in cluster i with luminosity greater than $L_\mu$ and less than $L_\mu + \Delta L_\mu$. 

The mean of this estimator can be written as  
\begin{eqnarray}
\label{eq:mean_old}
\langle\hat{\phi}(L_\mu, \lambda)\rangle = \langle\frac{\sum_i^{N_{cl}} N^g_i(L_\mu)}{N_{cl}(\lambda)\Delta L_\mu}\rangle_{g,P,s}, 
\end{eqnarray}
where in the above $\langle...\rangle_{g,P,s}$ denotes the average over the ensemble. Here we follow the formalism in \cite{2012MNRAS.426..531S}, which separated this process into three stages: g represents averaging over processes of populating galaxies into halos; p denotes averaging over processes of populating halos on a given dark matter density field; s represents averaging over gaussian sampling of the matter power spectrum within the survey volume. In other words, these three averaging processes represent averaging random processes of populating galaxies given a matter power spectrum under the assumption of the halo model. 

 Since we assume that the properties of galaxies only depend on the physical property of the host galaxy cluster,  we can simplify equation \ref{eq:mean_old} as 
\begin{eqnarray}
\label{eq:mean}
\langle\hat{\phi}(L_\mu, \lambda)\rangle &= & \langle\frac{\sum_i^{N_{cl}} \langle N^g_i(L_\mu)\rangle_g}{N_{cl}(\lambda)\Delta L_\mu}\rangle_{P,s} \nonumber\\
&= & \langle\frac{N^g (L_\mu)}{\Delta L_\mu}\rangle_{P,s}\nonumber \\
&= & \frac{N^g (L_\mu)}{\Delta L_\mu}.
\end{eqnarray}
Note that in the second line, we identify 
$\langle N^g_i(L_\mu)\rangle_g = N^g (L_\mu)$.

We then compute the covariance matrix of this estimator $C(\hat{\phi}(L_\mu, \lambda), \hat{\phi}(L_\nu, \lambda)) = \langle \hat{\phi}(L_\mu, \lambda) \hat{\phi}(L_\nu, \lambda)\rangle - \langle \hat{\phi}(L_\mu, \lambda)\rangle\langle \hat{\phi}(L_\nu, \lambda)\rangle $. 
First we focus on the first term. Inserting equation \ref{eq:estimator}, we have
\begin{equation}
\label{eq:first term}
 \langle \hat{\phi}(L_\mu) \hat{\phi}(L_\nu)\rangle = \langle\frac{\sum_i^{N_{cl}}\sum_j^{N_{cl}}\langle N_i^g(L_\mu)N_j^g(L_\nu)\rangle_g}{N_{cl}(\lambda)^2\Delta L_\mu \Delta L_\nu }\rangle_{P,s}
\end{equation}
Now, we focus on the inner bracket $\langle N_i^g(L_\mu)N_j^g(L_\mu)\rangle_g$. 
First, from our assumption that the properties of galaxies only depends on the physical property of the host galaxy cluster, we identify $\langle N_i^g(L_\mu)N_j^g(L_\nu)\rangle_g = \langle N_i^g(L_\mu)\rangle_g\langle N_j^g(L_\nu) \rangle_g$, if $i \neq j$. 
Also, under the assumption that each galaxy cluster has one and only one central, the term $i = j, \mu \neq \nu$ is zero. For the remaining term $i = j, \mu = \nu$, we identify
\begin{eqnarray}
\langle N_i^g(L_\mu)N_j^g(L_\mu)\rangle_g
&=&\langle N_i^g(L_\mu)^2\rangle_g\nonumber\\
&=&\langle N_i^g(L_\mu)\rangle_g\nonumber\\
&=&N^g(L_\mu),
\end{eqnarray} 
where in the second line, we adopt the assumption that the number of central galaxies in a galaxy cluster is one. In short, the inner bracket term can be written as 
\begin{eqnarray}
\label{eq:inner}
\langle N_i^g(L_\mu)N_j^g(L_\mu)\rangle_g = \epsilon_{i,j}N^g(L_\mu)N^g(L_\nu)+\delta_{i,j}\delta_{\mu,\nu}N^g(L_\mu),
\end{eqnarray}
where we use a modified Levi-Cevita symbol $\epsilon_{i,j}=1$ if $i\neq j$ and $0$ otherwise. 

Inserting equation \ref{eq:inner} back into equation \ref{eq:first term}, we have 
\begin{eqnarray}
\label{eq:final first}
\langle \hat{\phi}(L_\mu) \hat{\phi}(L_\nu)\rangle &=& \langle\frac{\sum_i^{N_{cl}}\sum_j^{N_{cl}}(\epsilon_{i,j}N^g(L_\mu)N^g(L_\nu)+\delta_{i,j}\delta_{\mu,\nu}N^g(L_\mu))}{N_{cl}(\lambda)^2\Delta L_\mu \Delta L_\nu }\rangle_{P,s}\nonumber\\
&=& \langle\frac{(N_{cl}(\lambda)^2-N_{cl}(\lambda))N^g(L_\mu)N^g(L_\nu)+N_{cl}(\lambda)\delta_{\mu,\nu}N^g(L_\mu)}{N_{cl}(\lambda)^2\Delta L_\mu \Delta L_\nu }\rangle_{P,s}\nonumber\\
&=& \frac{1}{\Delta L_\mu \Delta L_\nu}(\langle 1-\frac{1}{N_{cl}(
\lambda)}\rangle_{P,s}N^g(L_\mu)N^g(L_\nu) + \langle \frac{1}{N_{cl}(
\lambda)}\rangle_{P,s}N^g(L_\mu)\delta_{\mu, \nu})
\end{eqnarray}

Combining equation \ref{eq:mean} and equation \ref{eq:final first}, we obtain the covariance of central CLF estimator used in this paper, 
\begin{eqnarray}
C(\hat{\phi}(L_\mu, \lambda), \hat{\phi}(L_\nu, \lambda)) &=& \frac{1}{\Delta L_\mu \Delta L_\nu}(\langle -\frac{1}{N_{cl}(
\lambda)}\rangle_{P,s}N^g(L_\mu)N^g(L_\nu) + \langle \frac{1}{N_{cl}(
\lambda)}\rangle_{P,s}N^g(L_\mu)\delta_{\mu, \nu}). 
\end{eqnarray}
We further assume that $N_{cl}(\lambda)$ follows a Poisson distribution, which implies that $\langle 1/N_{cl}(\lambda)\rangle = 1/\langle N_{cl}(\lambda)\rangle$. We can further simplify our equation of covariance as
\begin{eqnarray}
\label{eq:final cov}
C(\hat{\phi}(L_\mu, \lambda), \hat{\phi}(L_\nu, \lambda)) 
&=& \frac{1}{\Delta L_\mu \Delta L_\nu}(-\frac{1}{\langle N_{cl}(
\lambda)\rangle_{P,s}}N^g(L_\mu)N^g(L_\nu) + \frac{1}{\langle N_{cl}(
\lambda)\rangle_{P,s}}N^g(L_\mu)\delta_{\mu, \nu})
\end{eqnarray}

\begin{figure}[ht!]
\includegraphics[width=0.7\textwidth]{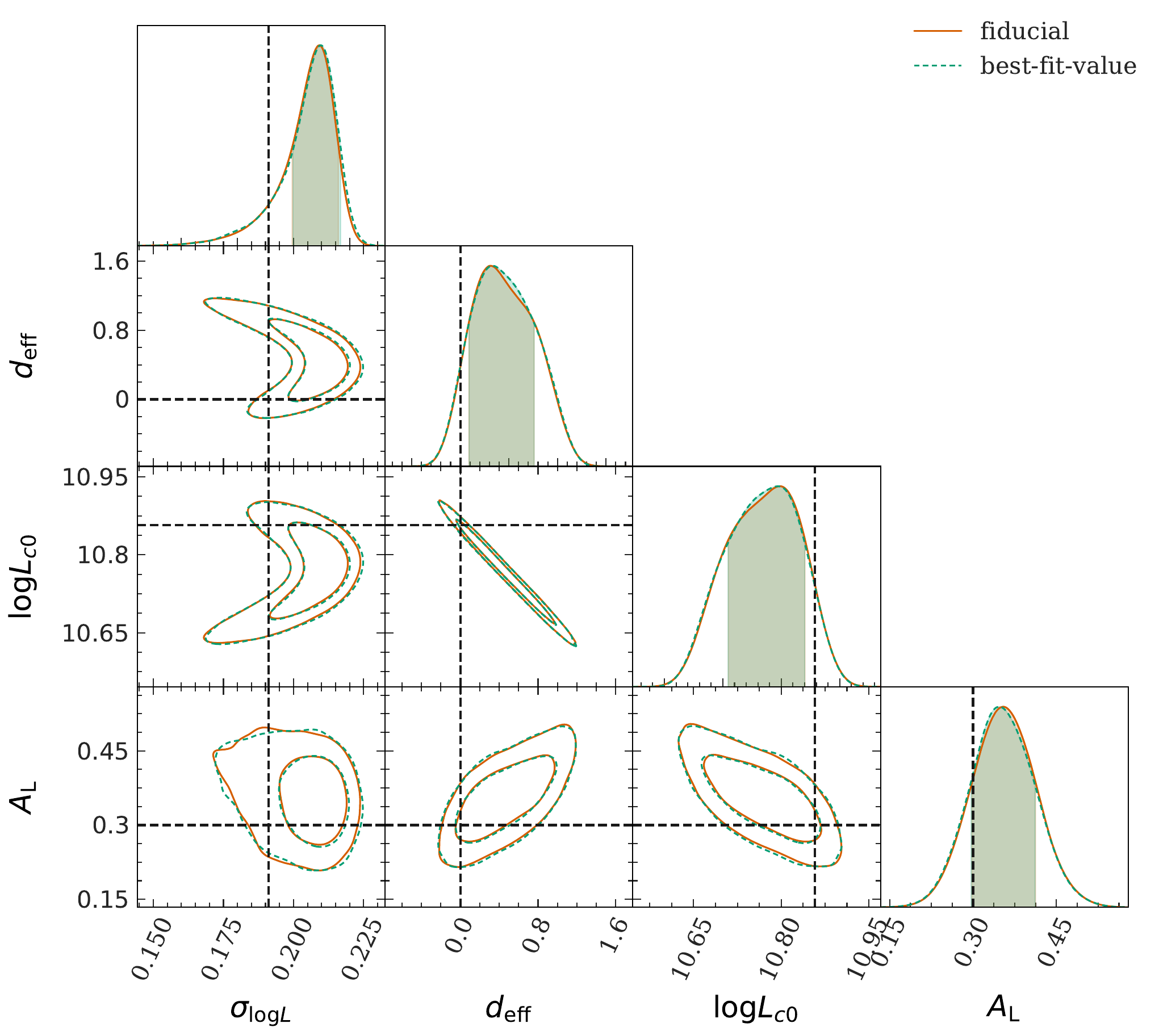}
\caption{Constraints on central CLF parameters using fiducial covarianc matrix (orange) and covariance matrix based on the CLF model centered at the maximum posterior (green dashed line). Here, we only perform this analysis on clusters in the first redshift bins of the main analysis, namely $z=0.1-0.15$. Since covariance matrices in different redshift bins are generated independently, an anlysis on a single redhsift bin is sufficient for this test. The dashed line indicates the fiducial parameters used to generate theoretical covariance. The agreement of constraints indicates that there is little dependence of the constraint on the choices of fiducial values used to generate the covariance matrix.} 
\label{fig-cov-fidicial}
\end{figure}

Following \cite{Hansen09}, we choose the parameter sets $\sigma_{\log L}, r, \log L_0, A_L = [0.44, 0, 25.0, 0.3]$ as our fiducial parameters to generate theoretical covariance matrices. We show in Figure \ref{fig-cov-fidicial} that the choice of fiducial values does not affect the posteriors at the accuracy of this analysis. 

\subsection{Covariance matrix form Jackknife resampling method}
\label{app-jackknifecovariance}
As a comparison, we construct an empirical covariance matrix from jackknife resampling method. We first divide the survey region into $N_{jk}$ simply connected patches using a k-means algorithm \footnote{\url{https://github.com/esheldon/kmeans_radec}}. We then remove one patch at a time and compute the conditional luminosity function of galaxies in the remaining patches. We use $\Phi_i$ to denote the luminosity function measured after removing the $i-th$ patch. The covariance matrix is given by

\begin{equation}
C =\frac{N_{jk}-1}{N_{jk}}\Sigma_{i}(\Phi_i-\bar{\Phi})^T(\Phi_i-\bar{\Phi}),
\end{equation}

where $\bar{\Phi}$ is the mean of $\Phi_i$. 
We choose $N_{jk}$ to be 150, thus each jackknife patch comprising $\sim 8\times8 \rm{deg}^2$ on the sky. At $z=0.1$, the lowest cluster redshift in this analysis, the jackknife region is $\sim 50\times 50 \rm{Mpc}^2$. We don't expect a significant correlation between clusters at this scale; therefore, each jackknife region could be considered independent. As a robustness check of the analysis, we verify that the result changes negligibly by varying $N_{jk}$ from 100 to 150. 

We notice that jackknife estimation is an noisy estimator of the covariance matrix, and the inverse of a noisy covariance matrix is a biased estimator of the inverse of covariance matrix. Various methods have been proposed to regularize the covariance matrix \citep{2008MNRAS.389..766P, 2015MNRAS.454.4326P, Olivier2018}. Here, because the number of jackknife regions is much larger than the number of entries in our data vector, we adopt a cut on the eigen-values of the covariance matrix. Specifically, we perform singular value decomposition on jackknife covariance matrix, and calculate the cumulative eigenvalues. We then truncate the last $0.5\%$ of the cumulative eigenvalues before we invert the covariance matrix.

\subsection{Covariance discussion}
\label{app:cov-disc}
\begin{figure}
\includegraphics[width=1.0\textwidth]{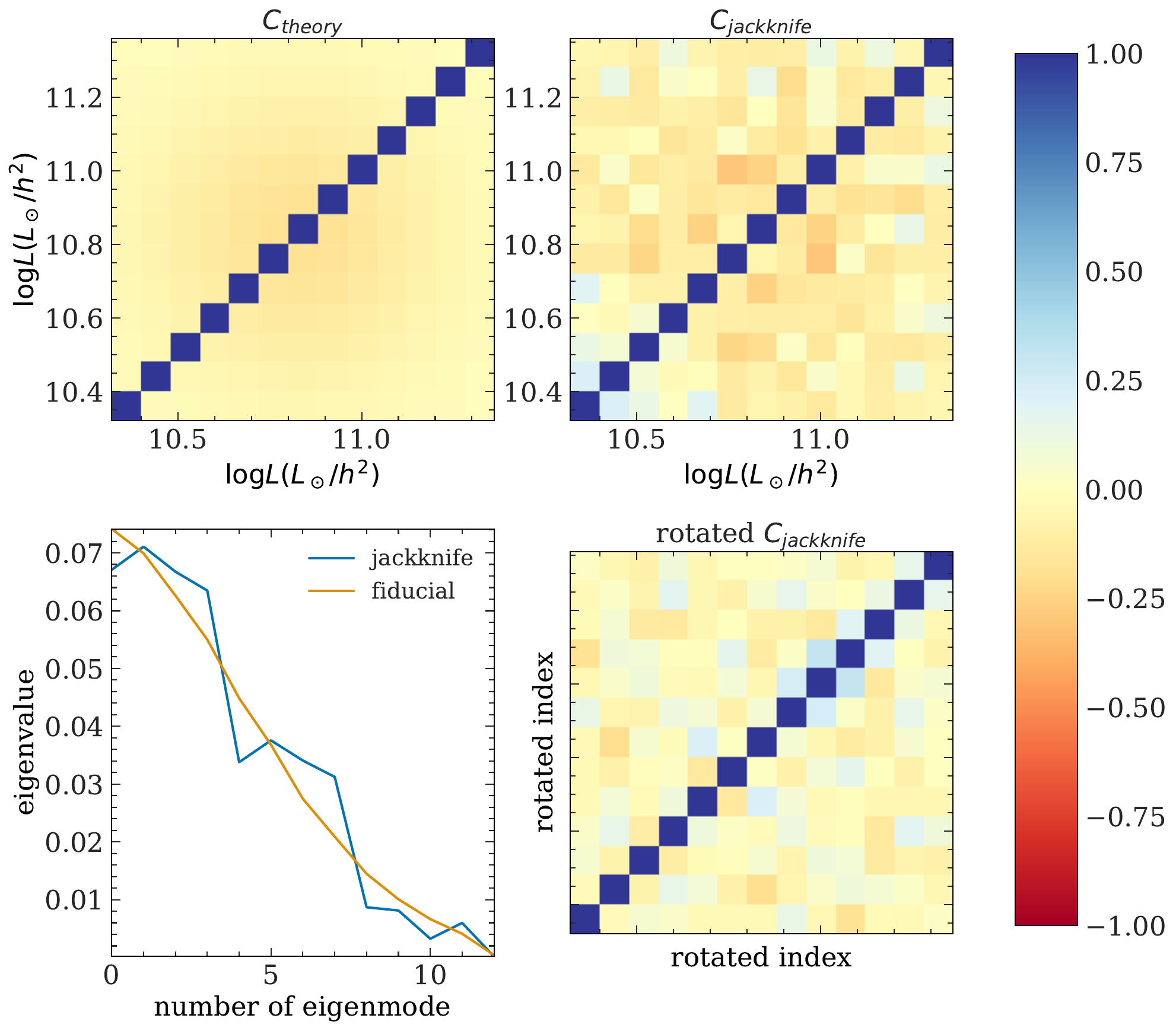}
\caption{Comparison between the theory covariance matrix and jackknife estimated covariance matrix. Top left: correlation matrix of the theory covariance matrix. Top right: correlation matrix of the jackknife covariance matrix. Bottom left: comparison of the eigenvalue of the theory covariance matrix and the diagonal term of the jackknife covariance matrix after rotated into the eigen-space of the theory covariance matrix. Bottom right: the correlation matrix of the jackknife covariance matrix after rotated into the eigen-space of the theory covariance matrix. See Appendix \ref{app:cov-disc} for more detailed descriptions}.
\label{fig-cov-comparison}
\end{figure}

\begin{figure}
\includegraphics[width=0.7\textwidth]{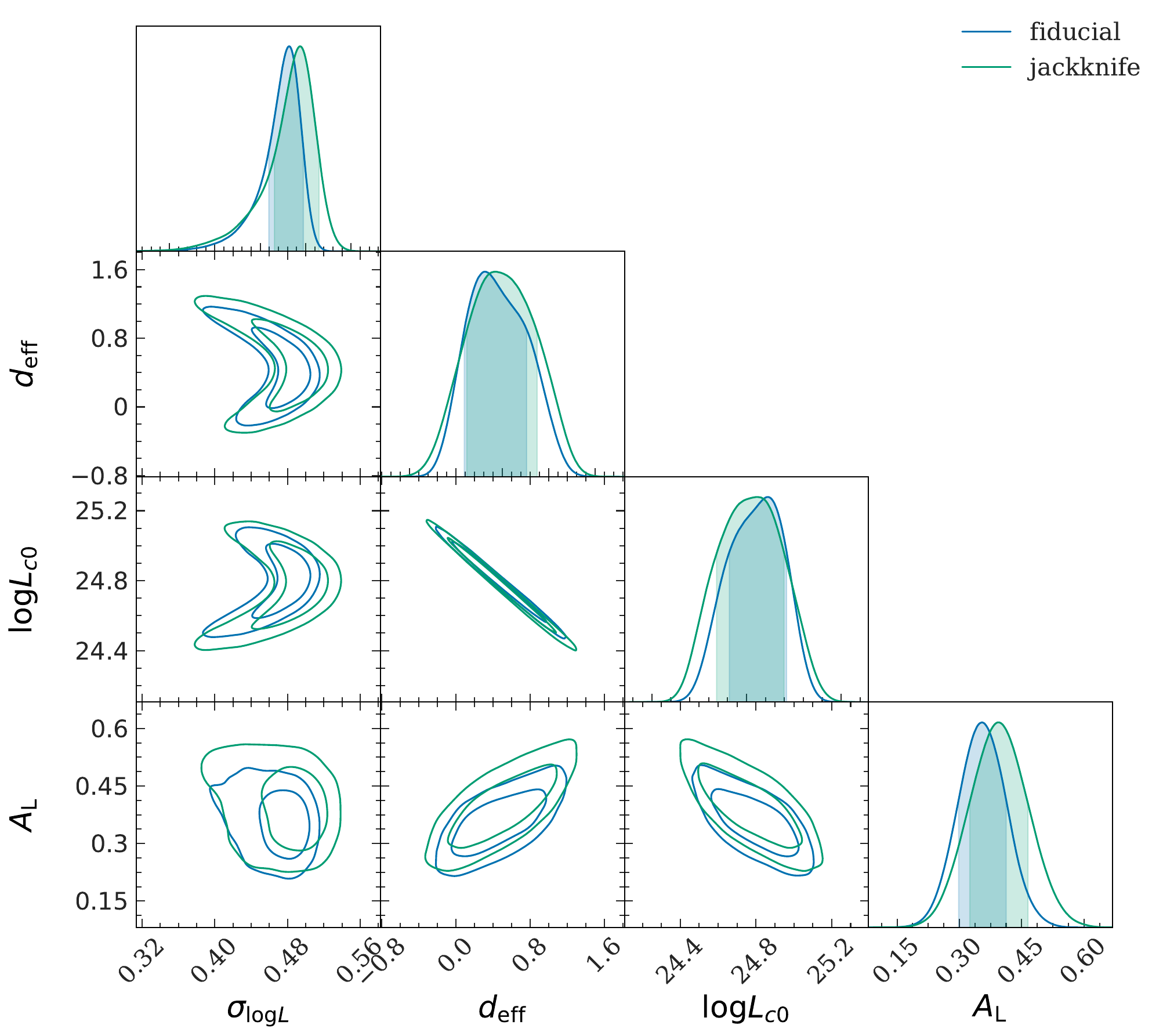}
\caption{$68\%$ and $95\%$ contours of the central CLF parameters. Both lines correspond to CLF data in the $z=[0.1,0.15]$ bin, but fitted with different covariance matrices: blue line represents constraints from using fiducial covariance matrix and the green line represents contraints from using empirical covariance matrix constructed using the process described in appendix \ref{app-jackknifecovariance}. The agreement of constraints indicates little dependence of the constraint on the choices of covariance matrix and justify the use of empirical covariance matrix for satellite CLFs.} 
\label{cov-jk-consistent}
\end{figure}

Figure \ref{fig-cov-comparison} demonstrates the difference between theoretical covariance matrix $C_{\rm{theory}}$ and jackknife estimated covariance matrix $C_{\rm{jackknife}}$. The top two panels show the correlation matrix of $C_{\rm{theory}}$ and $C_{\rm{jackknife}}$ respectively. It is hard to compare these two matrices directly, since there is noise in the jackknife covariance matrix. However, one could expect that the eigen-vectors with largest few eigen-values are less susceptible to the noise; hence we can make a better comparison of the covariance matrices by rotating $C_{\rm{jackknife}}$ into the eigen-space of $C_{\rm{theory}}$. 

If they are consistent, the rotated $C_{\rm{jackknife}}$ should be almost diagonal with the diagonal value similar to the eigen-vector of the theoretical covariance matrix. In the lower left panels of Figure \ref{fig-cov-comparison}, we show the comparison of the diagonal term of the rotated $C_{\rm{jackknife}}$ and the eigen-vector of the theoretical covariance matrix. The lower right panel shows the correlation matrix of the rotated $C_{\rm{jackknife}}$. We note that the eigen-value of $C_{\rm{theory}}$ and the diagonal term of the rotated $C_{\rm{jackknife}}$ are consistent. Moreover, the correlation matrix of the rotated $C_{\rm{jackknife}}$ is almost diagonal. Therefore, we conclude that $C_{\rm{theory}}$  and  $C_{\rm{jackknife}}$ are consistent with each other. 

Though we construct a reasonable theoretical covariance matrix for central CLFs, constructing the same thing for satellites is much harder. Constructing the covariance matrix for satellite CLF requires the knowledge of the correlation of satellites luminosity within the same host halos, which would require a much detailed model of halo galaxy connection and is beyond the scope of this paper. Instead, we use the empirical covariance matrix constructed using the same procedure as described in appendix \ref{app-jackknifecovariance}. To demonstrate the validity of using this covariance matrix, we show in Figure \ref{cov-jk-consistent} that the central CLF parameters obtained from this covariance matrix is consistent with the parameters obtained from the theoretical covariance matrix. Therefore, we expect the empirical covariance matrix not introducing a significant bias on our inferred satellite CLF parameters.  

\section{CLF of Individual Redshift Bins}

\begin{figure*}
\includegraphics[width=1.0\textwidth]{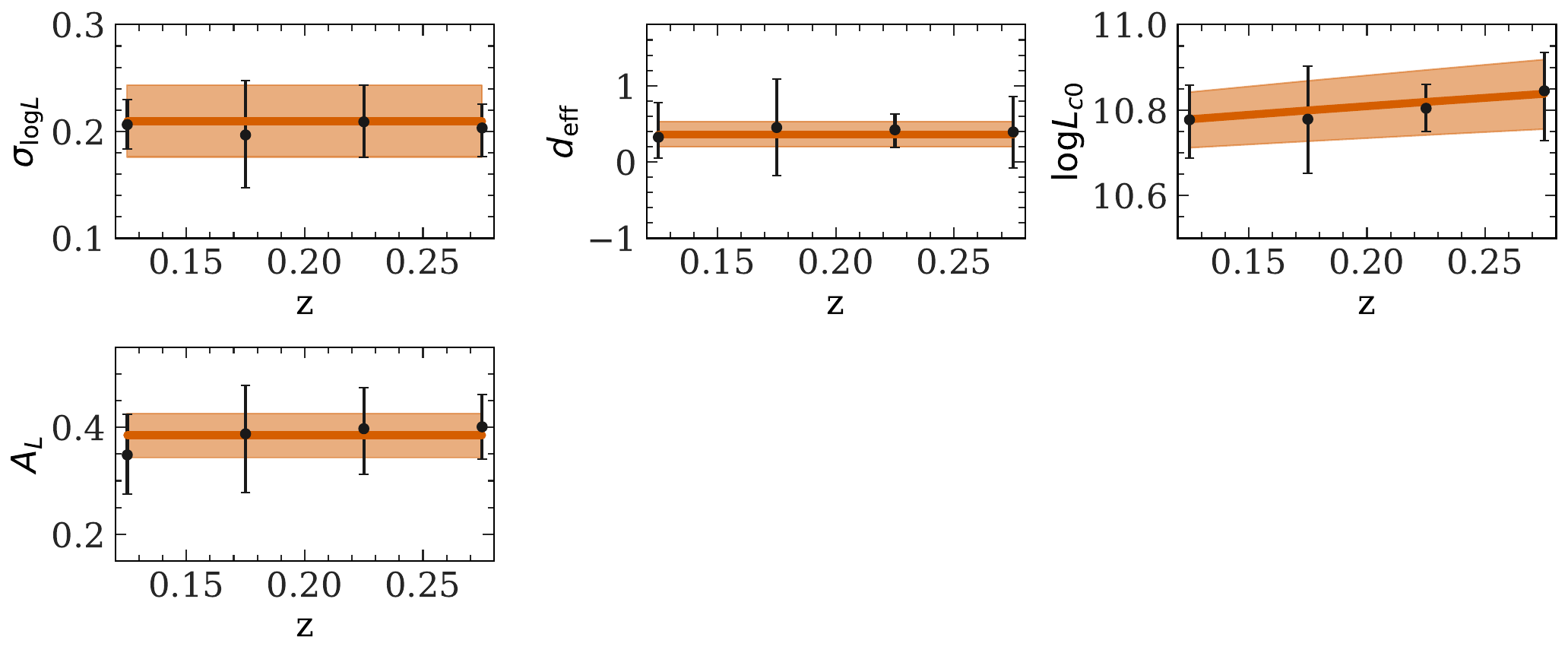}
\caption{Validation of redshift evolution of CLF model, for central galaxies.  Parameters fitted in individual redshift bins are shown with solid points and $1\sigma$ error bars including systematics. Solid lines are the results of CLF fits to all four rredshift bins simultaneously, including redshift evolution where appropriate. The color band represents one-sigma error including systematics.}
\label{fig-pev-cen}
\end{figure*}

\begin{figure*}
\includegraphics[width=1.0\textwidth]{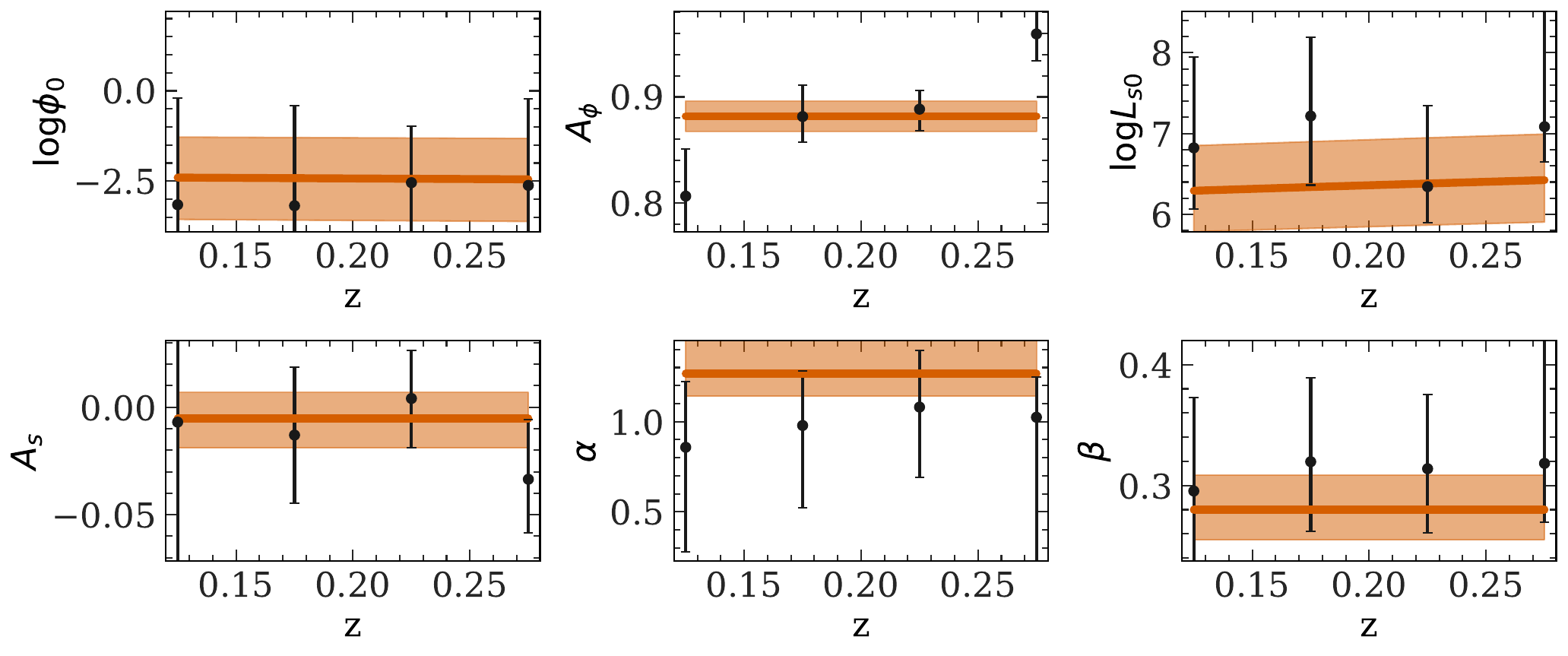}
\caption{Validation of redshift evolution of CLF model, for satellite galaxies.  Parameters fitted in individual redshift bins are shown with solid points and $1\sigma$ error bars including systematics. Solid lines are the results of CLF fit to all four redshift bins simultaneously, including redshift evolution where appropriate. The color band represents one-sigma error including systematics. Note that the $\log \phi_0$ panel only shows the parameter range allowed by the prior of the analysis.}
\label{fig-pev-sat}
\end{figure*}

We check whether our result is dominated by a single redshift bin and check unknown systematics by performing the fitting in each four redshift bin individually. In this analysis, we ignore the redshift evolution parameters for centrals $B_{\rm{L}}$ and satellites $B_{\rm{L}}$. The result is summarized in Table \ref{tab-clf-cen_ind} and Table \ref{tab-clf-sat_ind}. A comparison to the joint fit is shown in Figure \ref{fig-pev-cen} and Figure \ref{fig-pev-sat}.

\begingroup
\renewcommand{\arraystretch}{2.0}
\begin{table*}
    \centering
    \caption{Central Conditional Luminosity Function Parameters for individual redshift bins}
    \label{tab-clf-cen_ind}
    \begin{tabular}{ccccccc}
        \hline
		Redshift bins & $\sigma_{\log L}$ & $d_{\rm{eff}}$ & $logL_{0}$ & $A_L$  \\ 
        \hline
        Units & $\log L_\odot/h^2$ & - & $\log L_\odot/h^2$ & $\log L_\odot/h^2$ \\  
        
        Equation reference(s) &\ref{eq-cen-gen}& \ref{eq:chto_central} & \ref{eq-m-lcen} & \ref{eq-m-lcen}\\
		\hline
$0.1<z<0.15$ & $0.21^{+0.02}_{-0.02}$ & $0.32^{+0.45}_{-0.27}$ & $10.78^{+0.08}_{-0.09}$ & $0.35^{+0.08}_{-0.07}$\\
$0.15<z<0.2$ & $0.20^{+0.05}_{-0.05}$ & $0.45^{+0.64}_{-0.63}$ & $10.78^{+0.12}_{-0.13}$ & $0.39^{+0.09}_{-0.11}$\\
$0.2<z<0.25$ & $0.21^{+0.03}_{-0.03}$ & $0.42^{+0.21}_{-0.23}$ & $10.80^{+0.06}_{-0.05}$ & $0.40^{+0.08}_{-0.09}$\\
$0.25<z<0.3$ & $0.20^{+0.02}_{-0.03}$ & $0.39^{+0.46}_{-0.48}$ & $10.85^{+0.09}_{-0.12}$ & $0.40^{+0.06}_{-0.06}$\\

		\hline
    \end{tabular}
\end{table*}
\endgroup
\begingroup
\renewcommand{\arraystretch}{2.0}
\begin{table*}
    \centering
    \caption{Satellite Conditional Luminosity Function Parameters for individual redshift bins}
    \label{tab-clf-sat_ind}
    \begin{tabular}{ccccccccc}
        \hline
		Redshift bins & $\log\phi_0$ & $A_{\phi}$ & $\log L_{s0}$ & $A_s$ & $\alpha$ & $\beta$ \\ 
        \hline
        Units & $\log((\log L)^{-1}$ & $\log((\log L)^{-1}$ & $\log L_\odot/h^2$ & $\log L_\odot/h^2$ & - & -&\\  
        
        Equation reference(s) &\ref{eq-sat-gen},\ref{eq-phi}& \ref{eq-phi} & \ref{eq-sat-gen},\ref{eq-lst} & \ref{eq-lst} & \ref{eq-sat-gen}& \ref{eq-sat-gen}\\
		\hline
$0.1<z<0.15$ & $-3.15^{+2.96}_{-0.75}$ & $0.81^{+0.04}_{-0.03}$ & $6.82^{+1.13}_{-0.75}$ & $-0.01^{+0.04}_{-0.06}$ & $0.86^{+0.36}_{-0.58}$ & $0.30^{+0.08}_{-0.06}$\\
$0.15<z<0.2$ & $-3.18^{+2.77}_{-0.72}$ & $0.88^{+0.03}_{-0.02}$ & $7.22^{+0.98}_{-0.85}$ & $-0.01^{+0.03}_{-0.03}$ & $0.98^{+0.30}_{-0.45}$ & $0.32^{+0.07}_{-0.06}$\\
$0.2<z<0.25$ & $-2.54^{+1.56}_{-1.36}$ & $0.89^{+0.02}_{-0.02}$ & $6.34^{+1.00}_{-0.45}$ & $0.00^{+0.02}_{-0.02}$ & $1.08^{+0.32}_{-0.39}$ & $0.31^{+0.06}_{-0.05}$\\
$0.25<z<0.3$ & $-2.62^{+2.40}_{-1.28}$ & $0.96^{+0.02}_{-0.03}$ & $7.09^{+1.43}_{-0.44}$ & $-0.03^{+0.03}_{-0.02}$ & $1.02^{+0.22}_{-0.80}$ & $0.32^{+0.10}_{-0.05}$\\
		\hline
    \end{tabular}
\end{table*}
\endgroup

\clearpage
\newpage
\bibliography{references}{}

\begin{thebibliography}{}
\expandafter\ifx\csname natexlab\endcsname\relax\def\natexlab#1{#1}\fi

\bibitem[{{Abbott} {et~al.}(2018){Abbott}, {Abdalla}, {Alarcon}, {Aleksi{\'c}},
  {Allam}, {Allen}, {Amara}, {Annis}, {Asorey}, {Avila}, {Bacon}, {Balbinot},
  {Banerji}, {Banik}, {Barkhouse}, {Baumer}, {Baxter}, {Bechtol}, {Becker},
  {Benoit-L{\'e}vy}, {Benson}, {Bernstein}, {Bertin}, {Blazek}, {Bridle},
  {Brooks}, {Brout}, {Buckley-Geer}, {Burke}, {Busha}, {Campos}, {Capozzi},
  {Carnero Rosell}, {Carrasco Kind}, {Carretero}, {Castander}, {Cawthon},
  {Chang}, {Chen}, {Childress}, {Choi}, {Conselice}, {Crittenden}, {Crocce},
  {Cunha}, {D'Andrea}, {da Costa}, {Das}, {Davis}, {Davis}, {De Vicente},
  {DePoy}, {DeRose}, {Desai}, {Diehl}, {Dietrich}, {Dodelson}, {Doel},
  {Drlica-Wagner}, {Eifler}, {Elliott}, {Elsner}, {Elvin-Poole}, {Estrada},
  {Evrard}, {Fang}, {Fernandez}, {Fert{\'e}}, {Finley}, {Flaugher}, {Fosalba},
  {Friedrich}, {Frieman}, {Garc{\'{\i}}a-Bellido}, {Garcia-Fernandez}, {Gatti},
  {Gaztanaga}, {Gerdes}, {Giannantonio}, {Gill}, {Glazebrook}, {Goldstein},
  {Gruen}, {Gruendl}, {Gschwend}, {Gutierrez}, {Hamilton}, {Hartley}, {Hinton},
  {Honscheid}, {Hoyle}, {Huterer}, {Jain}, {James}, {Jarvis}, {Jeltema},
  {Johnson}, {Johnson}, {Kacprzak}, {Kent}, {Kim}, {King}, {Kirk}, {Kokron},
  {Kovacs}, {Krause}, {Krawiec}, {Kremin}, {Kuehn}, {Kuhlmann}, {Kuropatkin},
  {Lacasa}, {Lahav}, {Li}, {Liddle}, {Lidman}, {Lima}, {Lin}, {MacCrann},
  {Maia}, {Makler}, {Manera}, {March}, {Marshall}, {Martini}, {McMahon},
  {Melchior}, {Menanteau}, {Miquel}, {Miranda}, {Mudd}, {Muir}, {M{\"o}ller},
  {Neilsen}, {Nichol}, {Nord}, {Nugent}, {Ogando}, {Palmese}, {Peacock},
  {Peiris}, {Peoples}, {Percival}, {Petravick}, {Plazas}, {Porredon}, {Prat},
  {Pujol}, {Rau}, {Refregier}, {Ricker}, {Roe}, {Rollins}, {Romer}, {Roodman},
  {Rosenfeld}, {Ross}, {Rozo}, {Rykoff}, {Sako}, {Salvador}, {Samuroff},
  {S{\'a}nchez}, {Sanchez}, {Santiago}, {Scarpine}, {Schindler}, {Scolnic},
  {Secco}, {Serrano}, {Sevilla-Noarbe}, {Sheldon}, {Smith}, {Smith}, {Smith},
  {Soares-Santos}, {Sobreira}, {Suchyta}, {Tarle}, {Thomas}, {Troxel},
  {Tucker}, {Tucker}, {Uddin}, {Varga}, {Vielzeuf}, {Vikram}, {Vivas},
  {Walker}, {Wang}, {Wechsler}, {Weller}, {Wester}, {Wolf}, {Yanny}, {Yuan},
  {Zenteno}, {Zhang}, {Zhang}, {Zuntz}, \& {Dark Energy Survey
  Collaboration}}]{DESY1}
{Abbott}, T.~M.~C., {Abdalla}, F.~B., {Alarcon}, A., {et~al.} 2018, \prd, 98,
  043526

\bibitem[{{Aihara} {et~al.}(2011){Aihara}, {Allende Prieto}, {An}, {Anderson},
  {Aubourg}, {Balbinot}, {Beers}, {Berlind}, {Bickerton}, {Bizyaev}, {Blanton},
  {Bochanski}, {Bolton}, {Bovy}, {Brandt}, {Brinkmann}, {Brown}, {Brownstein},
  {Busca}, {Campbell}, {Carr}, {Chen}, {Chiappini}, {Comparat}, {Connolly},
  {Cortes}, {Croft}, {Cuesta}, {da Costa}, {Davenport}, {Dawson}, {Dhital},
  {Ealet}, {Ebelke}, {Edmondson}, {Eisenstein}, {Escoffier}, {Esposito},
  {Evans}, {Fan}, {Femen{\'{\i}}a Castell{\'a}}, {Font-Ribera}, {Frinchaboy},
  {Ge}, {Gillespie}, {Gilmore}, {Gonz{\'a}lez Hern{\'a}ndez}, {Gott}, {Gould},
  {Grebel}, {Gunn}, {Hamilton}, {Harding}, {Harris}, {Hawley}, {Hearty}, {Ho},
  {Hogg}, {Holtzman}, {Honscheid}, {Inada}, {Ivans}, {Jiang}, {Johnson},
  {Jordan}, {Jordan}, {Kazin}, {Kirkby}, {Klaene}, {Knapp}, {Kneib},
  {Kochanek}, {Koesterke}, {Kollmeier}, {Kron}, {Lampeitl}, {Lang}, {Le Goff},
  {Lee}, {Lin}, {Long}, {Loomis}, {Lucatello}, {Lundgren}, {Lupton}, {Ma},
  {MacDonald}, {Mahadevan}, {Maia}, {Makler}, {Malanushenko}, {Malanushenko},
  {Mandelbaum}, {Maraston}, {Margala}, {Masters}, {McBride}, {McGehee},
  {McGreer}, {M{\'e}nard}, {Miralda-Escud{\'e}}, {Morrison}, {Mullally},
  {Muna}, {Munn}, {Murayama}, {Myers}, {Naugle}, {Neto}, {Nguyen}, {Nichol},
  {O'Connell}, {Ogando}, {Olmstead}, {Oravetz}, {Padmanabhan},
  {Palanque-Delabrouille}, {Pan}, {Pandey}, {P{\^a}ris}, {Percival},
  {Petitjean}, {Pfaffenberger}, {Pforr}, {Phleps}, {Pichon}, {Pieri}, {Prada},
  {Price-Whelan}, {Raddick}, {Ramos}, {Reyl{\'e}}, {Rich}, {Richards}, {Rix},
  {Robin}, {Rocha-Pinto}, {Rockosi}, {Roe}, {Rollinde}, {Ross}, {Ross},
  {Rossetto}, {S{\'a}nchez}, {Sayres}, {Schlegel}, {Schlesinger}, {Schmidt},
  {Schneider}, {Sheldon}, {Shu}, {Simmerer}, {Simmons}, {Sivarani}, {Snedden},
  {Sobeck}, {Steinmetz}, {Strauss}, {Szalay}, {Tanaka}, {Thakar}, {Thomas},
  {Tinker}, {Tofflemire}, {Tojeiro}, {Tremonti}, {Vandenberg}, {Vargas
  Maga{\~n}a}, {Verde}, {Vogt}, {Wake}, {Wang}, {Weaver}, {Weinberg}, {White},
  {White}, {Yanny}, {Yasuda}, {Yeche}, \& {Zehavi}}]{SDSS2011}
{Aihara}, H., {Allende Prieto}, C., {An}, D., {et~al.} 2011, \apjs, 193, 29

\bibitem[{{Andrae} {et~al.}(2010){Andrae}, {Schulze-Hartung}, \&
  {Melchior}}]{2010arXiv1012.3754A}
{Andrae}, R., {Schulze-Hartung}, T., \& {Melchior}, P. 2010, ArXiv e-prints,
  arXiv:1012.3754

\bibitem[{{Behroozi} {et~al.}(2013){Behroozi}, {Wechsler}, \& {Wu}}]{Rockstar}
{Behroozi}, P.~S., {Wechsler}, R.~H., \& {Wu}, H.-Y. 2013, \apj, 762, 109

\bibitem[{{Bernardi} {et~al.}(2017){Bernardi}, {Fischer}, {Sheth}, {Meert},
  {Huertas-Company}, {Shankar}, \& {Vikram}}]{Ber17}
{Bernardi}, M., {Fischer}, J.-L., {Sheth}, R.~K., {et~al.} 2017, \mnras, 468,
  2569

\bibitem[{{Bernardi} {et~al.}(2013){Bernardi}, {Meert}, {Sheth}, {Vikram},
  {Huertas-Company}, {Mei}, \& {Shankar}}]{Ber13}
{Bernardi}, M., {Meert}, A., {Sheth}, R.~K., {et~al.} 2013, \mnras, 436, 697

\bibitem[{{Blanton} \& {Roweis}(2007)}]{BlRo2007}
{Blanton}, M.~R., \& {Roweis}, S. 2007, \aj, 133, 734

\bibitem[{{Blanton} {et~al.}(2005){Blanton}, {Schlegel}, {Strauss},
  {Brinkmann}, {Finkbeiner}, {Fukugita}, {Gunn}, {Hogg}, {Ivezi{\'c}}, {Knapp},
  {Lupton}, {Munn}, {Schneider}, {Tegmark}, \& {Zehavi}}]{Bla2005b}
{Blanton}, M.~R., {Schlegel}, D.~J., {Strauss}, M.~A., {et~al.} 2005, \aj, 129,
  2562

\bibitem[{{Budzynski} {et~al.}(2012){Budzynski}, {Koposov}, {McCarthy},
  {McGee}, \& {Belokurov}}]{Bud2012}
{Budzynski}, J.~M., {Koposov}, S.~E., {McCarthy}, I.~G., {McGee}, S.~L., \&
  {Belokurov}, V. 2012, \mnras, 423, 104

\bibitem[{{Cacciato} {et~al.}(2013){Cacciato}, {van den Bosch}, {More}, {Mo},
  \& {Yang}}]{Cacciato2013}
{Cacciato}, M., {van den Bosch}, F.~C., {More}, S., {Mo}, H., \& {Yang}, X.
  2013, \mnras, 430, 767

\bibitem[{{Chabrier}(2003)}]{Chabrier}
{Chabrier}, G. 2003, \pasp, 115, 763

\bibitem[{{Conroy} \& {Gunn}(2010)}]{SSP2}
{Conroy}, C., \& {Gunn}, J.~E. 2010, \apj, 712, 833

\bibitem[{{Conroy} {et~al.}(2009){Conroy}, {Gunn}, \& {White}}]{SSP1}
{Conroy}, C., {Gunn}, J.~E., \& {White}, M. 2009, \apj, 699, 486

\bibitem[{{Conroy} {et~al.}(2007){Conroy}, {Wechsler}, \& {Kravtsov}}]{CWK2007}
{Conroy}, C., {Wechsler}, R.~H., \& {Kravtsov}, A.~V. 2007, \apj, 668, 826

\bibitem[{{Costanzi} {et~al.}(2019{\natexlab{a}}){Costanzi}, {Rozo}, {Simet},
  {Zhang}, {Evrard}, {Mantz}, {Rykoff}, {Jeltema}, {Gruen}, {Allen},
  {McClintock}, {Romer}, {von der Linden}, {Farahi}, {DeRose}, {Varga},
  {Weller}, {Giles}, {Hollowood}, {Bhargava}, {Bermeo-Hernandez}, {Chen},
  {Abbott}, {Abdalla}, {Avila}, {Bechtol}, {Brooks}, {Buckley-Geer}, {Burke},
  {Rosell}, {Kind}, {Carretero}, {Crocce}, {Cunha}, {da Costa}, {Davis}, {De
  Vicente}, {Diehl}, {Dietrich}, {Doel}, {Eifler}, {Estrada}, {Flaugher},
  {Fosalba}, {Frieman}, {Garc{\'{\i}}a-Bellido}, {Gaztanaga}, {Gerdes},
  {Giannantonio}, {Gruendl}, {Gschwend}, {Gutierrez}, {Hartley}, {Honscheid},
  {Hoyle}, {James}, {Krause}, {Kuehn}, {Kuropatkin}, {Lima}, {Lin}, {Maia},
  {March}, {Marshall}, {Martini}, {Menanteau}, {Miller}, {Miquel}, {Mohr},
  {Ogando}, {Plazas}, {Roodman}, {Sanchez}, {Scarpine}, {Schindler},
  {Schubnell}, {Serrano}, {Sevilla-Noarbe}, {Sheldon}, {Smith},
  {Soares-Santos}, {Sobreira}, {Suchyta}, {Swanson}, {Tarle}, {Thomas}, \&
  {Wechsler}}]{SDSS_cluster_cosmology}
{Costanzi}, M., {Rozo}, E., {Simet}, M., {et~al.} 2019{\natexlab{a}}, \mnras,
  488, 4779

\bibitem[{{Costanzi} {et~al.}(2019{\natexlab{b}}){Costanzi}, {Rozo}, {Rykoff},
  {Farahi}, {Jeltema}, {Evrard}, {Mantz}, {Gruen}, {Mandelbaum}, {DeRose},
  {McClintock}, {Varga}, {Zhang}, {Weller}, {Wechsler}, \&
  {Aguena}}]{Projection_effect}
{Costanzi}, M., {Rozo}, E., {Rykoff}, E.~S., {et~al.} 2019{\natexlab{b}},
  \mnras, 482, 490

\bibitem[{{Deason} {et~al.}(2013){Deason}, {Conroy}, {Wetzel}, \&
  {Tinker}}]{Dea2013}
{Deason}, A.~J., {Conroy}, C., {Wetzel}, A.~R., \& {Tinker}, J.~L. 2013, \apj,
  777, 154

\bibitem[{{Diemer}(2018)}]{Colossus}
{Diemer}, B. 2018, \apjs, 239, 35

\bibitem[{{Diemer} {et~al.}(2013){Diemer}, {More}, \&
  {Kravtsov}}]{2013ApJ...766...25D}
{Diemer}, B., {More}, S., \& {Kravtsov}, A.~V. 2013, \apj, 766, 25

\bibitem[{{Driver} {et~al.}(2011){Driver}, {Hill}, {Kelvin}, {Robotham},
  {Liske}, {Norberg}, {Baldry}, {Bamford}, {Hopkins}, {Loveday}, {Peacock},
  {Andrae}, {Bland-Hawthorn}, {Brough}, {Brown}, {Cameron}, {Ching}, {Colless},
  {Conselice}, {Croom}, {Cross}, {de Propris}, {Dye}, {Drinkwater}, {Ellis},
  {Graham}, {Grootes}, {Gunawardhana}, {Jones}, {van Kampen}, {Maraston},
  {Nichol}, {Parkinson}, {Phillipps}, {Pimbblet}, {Popescu}, {Prescott},
  {Roseboom}, {Sadler}, {Sansom}, {Sharp}, {Smith}, {Taylor}, {Thomas},
  {Tuffs}, {Wijesinghe}, {Dunne}, {Frenk}, {Jarvis}, {Madore}, {Meyer},
  {Seibert}, {Staveley-Smith}, {Sutherland}, \& {Warren}}]{Dri2011}
{Driver}, S.~P., {Hill}, D.~T., {Kelvin}, L.~S., {et~al.} 2011, \mnras, 413,
  971

\bibitem[{{Foreman-Mackey} {et~al.}(2013){Foreman-Mackey}, {Hogg}, {Lang}, \&
  {Goodman}}]{For2012}
{Foreman-Mackey}, D., {Hogg}, D.~W., {Lang}, D., \& {Goodman}, J. 2013, \pasp,
  125, 306

\bibitem[{{Friedrich} \& {Eifler}(2018)}]{Olivier2018}
{Friedrich}, O., \& {Eifler}, T. 2018, \mnras, 473, 4150

\bibitem[{{Hansen} {et~al.}(2009){Hansen}, {Sheldon}, {Wechsler}, \&
  {Koester}}]{Hansen09}
{Hansen}, S.~M., {Sheldon}, E.~S., {Wechsler}, R.~H., \& {Koester}, B.~P. 2009,
  \apj, 699, 1333

\bibitem[{{Hearin} {et~al.}(2013){Hearin}, {Zentner}, {Newman}, \&
  {Berlind}}]{Hea2012}
{Hearin}, A.~P., {Zentner}, A.~R., {Newman}, J.~A., \& {Berlind}, A.~A. 2013,
  \mnras, 430, 1238

\bibitem[{Hoshino {et~al.}(2015)Hoshino, Leauthaud, More, Vulcani, Lackner,
  Saito, More, Hikage, Rozo, Rykoff, \& Mandelbaum}]{Hoshino2015}
Hoshino, H., Leauthaud, A., More, A., {et~al.} 2015, Monthly Notices of the
  Royal Astronomical Society, 452, 998

\bibitem[{Hunter(2007)}]{Hunter:2007}
Hunter, J.~D. 2007, Computing in Science \& Engineering, 9, 90

\bibitem[{{Klypin} {et~al.}(2016){Klypin}, {Yepes}, {Gottl{\"o}ber}, {Prada},
  \& {He{\ss}}}]{MDPL}
{Klypin}, A., {Yepes}, G., {Gottl{\"o}ber}, S., {Prada}, F., \& {He{\ss}}, S.
  2016, \mnras, 457, 4340

\bibitem[{{Kravtsov} {et~al.}(2018){Kravtsov}, {Vikhlinin}, \&
  {Meshcheryakov}}]{Kravstov2018}
{Kravtsov}, A.~V., {Vikhlinin}, A.~A., \& {Meshcheryakov}, A.~V. 2018,
  Astronomy Letters, 44, 8

\bibitem[{{Kunth} \& {{\"O}stlin}(2000)}]{2000A&ARv..10....1K}
{Kunth}, D., \& {{\"O}stlin}, G. 2000, \aapr, 10, 1

\bibitem[{{Lange} {et~al.}(2018){Lange}, {van den Bosch}, {Hearin}, {Campbell},
  {Zentner}, {Villarreal}, \& {Mao}}]{Lange2018}
{Lange}, J.~U., {van den Bosch}, F.~C., {Hearin}, A., {et~al.} 2018, \mnras,
  473, 2830

\bibitem[{{Leauthaud} {et~al.}(2012){Leauthaud}, {Tinker}, {Bundy}, {Behroozi},
  {Massey}, {Rhodes}, {George}, {Kneib}, {Benson}, {Wechsler}, {Busha},
  {Capak}, {Cort{\^e}s}, {Ilbert}, {Koekemoer}, {Le F{\`e}vre}, {Lilly},
  {McCracken}, {Salvato}, {Schrabback}, {Scoville}, {Smith}, \&
  {Taylor}}]{2012ApJ...744..159L}
{Leauthaud}, A., {Tinker}, J., {Bundy}, K., {et~al.} 2012, \apj, 744, 159

\bibitem[{{Lehmann} {et~al.}(2017){Lehmann}, {Mao}, {Becker}, {Skillman}, \&
  {Wechsler}}]{Lehmann2017}
{Lehmann}, B.~V., {Mao}, Y.-Y., {Becker}, M.~R., {Skillman}, S.~W., \&
  {Wechsler}, R.~H. 2017, \apj, 834, 37

\bibitem[{{Li} {et~al.}(2014){Li}, {Shan}, {Mo}, {Kneib}, {Yang}, {Luo}, {van
  den Bosch}, {Erben}, {Moraes}, \& {Makler}}]{2014MNRAS.438.2864L}
{Li}, R., {Shan}, H., {Mo}, H., {et~al.} 2014, Monthly Notices of the Royal
  Astronomical Society, 438, 2864

\bibitem[{{Lin} {et~al.}(2013){Lin}, {Brodwin}, {Gonzalez}, {Bode},
  {Eisenhardt}, {Stanford}, \& {Vikhlinin}}]{Lin2013}
{Lin}, Y.-T., {Brodwin}, M., {Gonzalez}, A.~H., {et~al.} 2013, \apj, 771, 61

\bibitem[{{Lin} {et~al.}(2004){Lin}, {Mohr}, \& {Stanford}}]{LMS2004}
{Lin}, Y.-T., {Mohr}, J.~J., \& {Stanford}, S.~A. 2004, \apj, 610, 745

\bibitem[{{Lin} {et~al.}(2010){Lin}, {Ostriker}, \& {Miller}}]{LOM2010}
{Lin}, Y.-T., {Ostriker}, J.~P., \& {Miller}, C.~J. 2010, \apj, 715, 1486

\bibitem[{{Mancone} \& {Gonzalez}(2012)}]{EZgal}
{Mancone}, C.~L., \& {Gonzalez}, A.~H. 2012, \pasp, 124, 606

\bibitem[{{Mao} {et~al.}(2015){Mao}, {Williamson}, \& {Wechsler}}]{mao2015}
{Mao}, Y.-Y., {Williamson}, M., \& {Wechsler}, R.~H. 2015, \apj, 810, 21

\bibitem[{{Meert} {et~al.}(2015){Meert}, {Vikram}, \& {Bernardi}}]{Meert2015}
{Meert}, A., {Vikram}, V., \& {Bernardi}, M. 2015, \mnras, 446, 3943

\bibitem[{{Meert} {et~al.}(2016){Meert}, {Vikram}, \& {Bernardi}}]{Meert2016}
---. 2016, \mnras, 455, 2440

\bibitem[{{More}(2012)}]{More2012}
{More}, S. 2012, \apj, 761, 127

\bibitem[{{Murray} {et~al.}(2013){Murray}, {Power}, \& {Robotham}}]{MPR2013}
{Murray}, S.~G., {Power}, C., \& {Robotham}, A.~S.~G. 2013, Astronomy and
  Computing, 3, 23

\bibitem[{{Paranjape} \& {Sheth}(2012)}]{PS2012}
{Paranjape}, A., \& {Sheth}, R.~K. 2012, \mnras, 423, 1845

\bibitem[{{Paz} \& {S{\'a}nchez}(2015)}]{2015MNRAS.454.4326P}
{Paz}, D.~J., \& {S{\'a}nchez}, A.~G. 2015, \mnras, 454, 4326

\bibitem[{{Peng} {et~al.}(2010){Peng}, {Lilly}, {Kova{\v{c}}}, {Bolzonella},
  {Pozzetti}, {Renzini}, {Zamorani}, {Ilbert}, {Knobel}, {Iovino}, {Maier},
  {Cucciati}, {Tasca}, {Carollo}, {Silverman}, {Kampczyk}, {de Ravel},
  {Sanders}, {Scoville}, {Contini}, {Mainieri}, {Scodeggio}, {Kneib}, {Le
  F{\`e}vre}, {Bardelli}, {Bongiorno}, {Caputi}, {Coppa}, {de la Torre},
  {Franzetti}, {Garilli}, {Lamareille}, {Le Borgne}, {Le Brun}, {Mignoli},
  {Perez Montero}, {Pello}, {Ricciardelli}, {Tanaka}, {Tresse}, {Vergani},
  {Welikala}, {Zucca}, {Oesch}, {Abbas}, {Barnes}, {Bordoloi}, {Bottini},
  {Cappi}, {Cassata}, {Cimatti}, {Fumana}, {Hasinger}, {Koekemoer},
  {Leauthaud}, {Maccagni}, {Marinoni}, {McCracken}, {Memeo}, {Meneux}, {Nair},
  {Porciani}, {Presotto}, \& {Scaramella}}]{2010ApJ...721..193P}
{Peng}, Y.-j., {Lilly}, S.~J., {Kova{\v{c}}}, K., {et~al.} 2010, \apj, 721, 193

\bibitem[{{Planck Collaboration} {et~al.}(2014){Planck Collaboration}, {Ade},
  {Aghanim}, {Armitage-Caplan}, {Arnaud}, {Ashdown}, {Atrio-Barandela},
  {Aumont}, {Baccigalupi}, {Banday}, \& et~al.}]{Planck13}
{Planck Collaboration}, {Ade}, P.~A.~R., {Aghanim}, N., {et~al.} 2014, \aap,
  571, A16

\bibitem[{{Pope} \& {Szapudi}(2008)}]{2008MNRAS.389..766P}
{Pope}, A.~C., \& {Szapudi}, I. 2008, \mnras, 389, 766

\bibitem[{{Reddick} {et~al.}(2013){Reddick}, {Wechsler}, {Tinker}, \&
  {Behroozi}}]{Reddick2013}
{Reddick}, R.~M., {Wechsler}, R.~H., {Tinker}, J.~L., \& {Behroozi}, P.~S.
  2013, \apj, 771, 30

\bibitem[{{Rozo} \& {Rykoff}(2014)}]{RoRy2013}
{Rozo}, E., \& {Rykoff}, E.~S. 2014, \apj, 783, 80

\bibitem[{{Rozo} {et~al.}(2015{\natexlab{a}}){Rozo}, {Rykoff}, {Becker},
  {Reddick}, \& {Wechsler}}]{Rozo2014}
{Rozo}, E., {Rykoff}, E.~S., {Becker}, M., {Reddick}, R.~M., \& {Wechsler},
  R.~H. 2015{\natexlab{a}}, \mnras, 453, 38

\bibitem[{{Rozo} {et~al.}(2015{\natexlab{b}}){Rozo}, {Rykoff}, {Becker},
  {Reddick}, \& {Wechsler}}]{Rozo2015}
---. 2015{\natexlab{b}}, \mnras, 453, 38

\bibitem[{{Rozo} {et~al.}(2009){Rozo}, {Rykoff}, {Evrard}, {Becker}, {McKay},
  {Wechsler}, {Koester}, {Hao}, {Hansen}, {Sheldon}, {Johnston}, {Annis}, \&
  {Frieman}}]{Rozo09}
{Rozo}, E., {Rykoff}, E.~S., {Evrard}, A., {et~al.} 2009, \apj, 699, 768

\bibitem[{{Rykoff} {et~al.}(2014{\natexlab{a}}){Rykoff}, {Rozo}, {Busha},
  {Cunha}, {Finoguenov}, {Evrard}, {Hao}, {Koester}, {Leauthaud}, {Nord},
  {Pierre}, {Reddick}, {Sadibekova}, {Sheldon}, \& {Wechsler}}]{Redmapper1}
{Rykoff}, E.~S., {Rozo}, E., {Busha}, M.~T., {et~al.} 2014{\natexlab{a}}, \apj,
  785, 104

\bibitem[{{Rykoff} {et~al.}(2014{\natexlab{b}}){Rykoff}, {Rozo}, {Busha},
  {Cunha}, {Finoguenov}, {Evrard}, {Hao}, {Koester}, {Leauthaud}, {Nord},
  {Pierre}, {Reddick}, {Sadibekova}, {Sheldon}, \& {Wechsler}}]{Ryk2013}
---. 2014{\natexlab{b}}, \apj, 785, 104

\bibitem[{{Shan} {et~al.}(2015){Shan}, {McDonald}, \& {Courteau}}]{Masstolight}
{Shan}, Y., {McDonald}, M., \& {Courteau}, S. 2015, \apj, 800, 122

\bibitem[{{Simet} {et~al.}(2017){Simet}, {McClintock}, {Mandelbaum}, {Rozo},
  {Rykoff}, {Sheldon}, \& {Wechsler}}]{simetetal17}
{Simet}, M., {McClintock}, T., {Mandelbaum}, R., {et~al.} 2017, \mnras, 466,
  3103

\bibitem[{{Skibba} {et~al.}(2011){Skibba}, {van den Bosch}, {Yang}, {More},
  {Mo}, \& {Fontanot}}]{Ski2011}
{Skibba}, R.~A., {van den Bosch}, F.~C., {Yang}, X., {et~al.} 2011, \mnras,
  410, 417

\bibitem[{{Smith}(2012{\natexlab{a}})}]{covariance}
{Smith}, R.~E. 2012{\natexlab{a}}, \mnras, 426, 531

\bibitem[{{Smith}(2012{\natexlab{b}})}]{2012MNRAS.426..531S}
---. 2012{\natexlab{b}}, \mnras, 426, 531

\bibitem[{{Tal} {et~al.}(2013){Tal}, {van Dokkum}, {Franx}, {Leja}, {Wake}, \&
  {Whitaker}}]{Tal2013}
{Tal}, T., {van Dokkum}, P.~G., {Franx}, M., {et~al.} 2013, \apj, 769, 31

\bibitem[{{Tavasoli} {et~al.}(2011){Tavasoli}, {Khosroshahi}, {Koohpaee},
  {Rahmani}, \& {Ghanbari}}]{Tav2011}
{Tavasoli}, S., {Khosroshahi}, H.~G., {Koohpaee}, A., {Rahmani}, H., \&
  {Ghanbari}, J. 2011, \pasp, 123, 1

\bibitem[{{Tinker} {et~al.}(2008){Tinker}, {Kravtsov}, {Klypin}, {Abazajian},
  {Warren}, {Yepes}, {Gottl{\"o}ber}, \& {Holz}}]{Tinker2008}
{Tinker}, J., {Kravtsov}, A.~V., {Klypin}, A., {et~al.} 2008, \apj, 688, 709

\bibitem[{{Tollet} {et~al.}(2017){Tollet}, {Cattaneo}, {Mamon}, {Moutard}, \&
  {van den Bosch}}]{Tidalstriping}
{Tollet}, {\'E}., {Cattaneo}, A., {Mamon}, G.~A., {Moutard}, T., \& {van den
  Bosch}, F.~C. 2017, \mnras, 471, 4170

\bibitem[{{van den Bosch} {et~al.}(2008){van den Bosch}, {Aquino}, {Yang},
  {Mo}, {Pasquali}, {McIntosh}, {Weinmann}, \& {Kang}}]{2008MNRAS.387...79V}
{van den Bosch}, F.~C., {Aquino}, D., {Yang}, X., {et~al.} 2008, \mnras, 387,
  79

\bibitem[{{van der Walt} {et~al.}(2011){van der Walt}, {Colbert}, \&
  {Varoquaux}}]{numpy}
{van der Walt}, S., {Colbert}, S.~C., \& {Varoquaux}, G. 2011, Computing in
  Science Engineering, 13, 22

\bibitem[{{Watson} {et~al.}(2012){Watson}, {Berlind}, \& {Zentner}}]{WBZ2012}
{Watson}, D.~F., {Berlind}, A.~A., \& {Zentner}, A.~R. 2012, \apj, 754, 90

\bibitem[{{Wechsler} \& {Tinker}(2018)}]{RisaAwesomepaper}
{Wechsler}, R.~H., \& {Tinker}, J.~L. 2018, \araa, 56, 435

\bibitem[{{Wetzel} {et~al.}(2012){Wetzel}, {Tinker}, \& {Conroy}}]{WTC2011}
{Wetzel}, A.~R., {Tinker}, J.~L., \& {Conroy}, C. 2012, \mnras, 424, 232

\bibitem[{{Wetzel} {et~al.}(2013){Wetzel}, {Tinker}, {Conroy}, \& {van den
  Bosch}}]{Wet13}
{Wetzel}, A.~R., {Tinker}, J.~L., {Conroy}, C., \& {van den Bosch}, F.~C. 2013,
  \mnras, 432, 336

\bibitem[{{Wu} {et~al.}(2013){Wu}, {Hahn}, {Wechsler}, {Behroozi}, \&
  {Mao}}]{Wu2013}
{Wu}, H.-Y., {Hahn}, O., {Wechsler}, R.~H., {Behroozi}, P.~S., \& {Mao}, Y.-Y.
  2013, \apj, 767, 23

\bibitem[{{Yang} {et~al.}(2008){Yang}, {Mo}, \& {van den Bosch}}]{YMB2008}
{Yang}, X., {Mo}, H.~J., \& {van den Bosch}, F.~C. 2008, \apj, 676, 248

\bibitem[{{Yang} {et~al.}(2009){Yang}, {Mo}, \& {van den Bosch}}]{YMB2009}
---. 2009, \apj, 695, 900

\bibitem[{{Yang} {et~al.}(2005){Yang}, {Mo}, {van den Bosch}, \&
  {Jing}}]{YMB2005}
{Yang}, X., {Mo}, H.~J., {van den Bosch}, F.~C., \& {Jing}, Y.~P. 2005, \mnras,
  356, 1293

\bibitem[{{Zhang} {et~al.}(2016){Zhang}, {Miller}, {McKay}, {Rooney}, {Evrard},
  {Romer}, {Perfecto}, {Song}, {Desai}, {Mohr}, {Wilcox}, {Bermeo-Hernandez},
  {Jeltema}, {Hollowood}, {Bacon}, {Capozzi}, {Collins}, {Das}, {Gerdes},
  {Hennig}, {Hilton}, {Hoyle}, {Kay}, {Liddle}, {Mann}, {Mehrtens}, {Nichol},
  {Papovich}, {Sahl{\'e}n}, {Soares-Santos}, {Stott}, {Viana}, {Abbott},
  {Abdalla}, {Banerji}, {Bauer}, {Benoit-L{\'e}vy}, {Bertin}, {Brooks},
  {Buckley-Geer}, {Burke}, {Carnero Rosell}, {Castander}, {Diehl}, {Doel},
  {Cunha}, {Eifler}, {Fausti Neto}, {Fernandez}, {Flaugher}, {Fosalba},
  {Frieman}, {Gaztanaga}, {Gruen}, {Gruendl}, {Honscheid}, {James}, {Kuehn},
  {Kuropatkin}, {Lahav}, {Maia}, {Makler}, {Marshall}, {Martini}, {Miquel},
  {Ogando}, {Plazas}, {Roodman}, {Rykoff}, {Sako}, {Sanchez}, {Scarpine},
  {Schubnell}, {Sevilla}, {Smith}, {Sobreira}, {Suchyta}, {Swanson}, {Tarle},
  {Thaler}, {Tucker}, {Vikram}, \& {da Costa}}]{Zhang2016}
{Zhang}, Y., {Miller}, C., {McKay}, T., {et~al.} 2016, \apj, 816, 98

\bibitem[{{Zhang} {et~al.}(2019{\natexlab{a}}){Zhang}, {Jeltema}, {Hollowood},
  {Everett}, {Rozo}, {Farahi}, {Bermeo}, {Bhargava}, {Giles}, {Romer},
  {Wilkinson}, {Rykoff}, {Mantz}, {Diehl}, {Evrard}, {Stern}, {Gruen}, {von der
  Linden}, {Splettstoesser}, {Chen}, {Costanzi}, {Allen}, {Collins}, {Hilton},
  {Klein}, {Mann}, {Manolopoulou}, {Morris}, {Mayers}, {Sahlen}, {Stott},
  {Vergara Cervantes}, {Viana}, {Wechsler}, {Allam}, {Avila}, {Bechtol},
  {Bertin}, {Brooks}, {Burke}, {Carnero Rosell}, {Carrasco Kind}, {Carretero},
  {Castander}, {da Costa}, {De Vicente}, {Desai}, {Dietrich}, {Doel},
  {Flaugher}, {Fosalba}, {Frieman}, {Garc{\'{\i}}a-Bellido}, {Gaztanaga},
  {Gruendl}, {Gschwend}, {Gutierrez}, {Hartley}, {Honscheid}, {Hoyle},
  {Krause}, {Kuehn}, {Kuropatkin}, {Lima}, {Maia}, {Marshall}, {Melchior},
  {Menanteau}, {Miller}, {Miquel}, {Ogando}, {Plazas}, {Sanchez}, {Scarpine},
  {Schindler}, {Serrano}, {Sevilla-Noarbe}, {Smith}, {Soares-Santos},
  {Suchyta}, {Swanson}, {Tarle}, {Thomas}, {Tucker}, {Vikram}, \&
  {Wester}}]{Miscentering}
{Zhang}, Y., {Jeltema}, T., {Hollowood}, D.~L., {et~al.} 2019{\natexlab{a}},
  arXiv e-prints, arXiv:1901.07119

\bibitem[{{Zhang} {et~al.}(2019{\natexlab{b}}){Zhang}, {Yanny}, {Palmese},
  {Gruen}, {To}, {Rykoff}, {Leung}, {Collins}, {Hilton}, {Abbott}, {Annis},
  {Avila}, {Bertin}, {Brooks}, {Burke}, {Carnero Rosell}, {Carrasco Kind},
  {Carretero}, {Cunha}, {D'Andrea}, {da Costa}, {De Vicente}, {Desai}, {Diehl},
  {Dietrich}, {Doel}, {Drlica-Wagner}, {Eifler}, {Evrard}, {Flaugher},
  {Fosalba}, {Frieman}, {Garc{\'{\i}}a-Bellido}, {Gaztanaga}, {Gerdes},
  {Gruendl}, {Gschwend}, {Gutierrez}, {Hartley}, {Hollowood}, {Honscheid},
  {Hoyle}, {James}, {Jeltema}, {Kuehn}, {Kuropatkin}, {Li}, {Lima}, {Maia},
  {March}, {Marshall}, {Melchior}, {Menanteau}, {Miller}, {Miquel}, {Mohr},
  {Ogando}, {Plazas}, {Romer}, {Sanchez}, {Scarpine}, {Schubnell}, {Serrano},
  {Sevilla-Noarbe}, {Smith}, {Soares-Santos}, {Sobreira}, {Suchyta}, {Swanson},
  {Tarle}, {Thomas}, {Wester}, \& {DES Collaboration}}]{Zhang18}
{Zhang}, Y., {Yanny}, B., {Palmese}, A., {et~al.} 2019{\natexlab{b}}, \apj,
  874, 165

\bibitem[{{Zhang} {et~al.}(2019{\natexlab{c}}){Zhang}, {Miller}, {Rooney},
  {Bermeo}, {Romer}, {Vergara Cervantes}, {Rykoff}, {Hennig}, {Das}, {McKay},
  {Song}, {Wilcox}, {Bacon}, {Bridle}, {Collins}, {Conselice}, {Hilton},
  {Hoyle}, {Kay}, {Liddle}, {Mann}, {Mehrtens}, {Mayers}, {Nichol},
  {Sahl{\'e}n}, {Stott}, {Viana}, {Wechsler}, {Abbott}, {Abdalla}, {Allam},
  {Benoit-L{\'e}vy}, {Brooks}, {Buckley-Geer}, {Burke}, {Carnero Rosell},
  {Carrasco Kind}, {Carretero}, {Castander}, {Crocce}, {Cunha}, {D'Andrea}, {da
  Costa}, {Diehl}, {Dietrich}, {Eifler}, {Flaugher}, {Fosalba},
  {Garc{\'{\i}}a-Bellido}, {Gaztanaga}, {Gerdes}, {Gruen}, {Gruendl},
  {Gschwend}, {Gutierrez}, {Honscheid}, {James}, {Jeltema}, {Kuehn},
  {Kuropatkin}, {Lima}, {Lin}, {Maia}, {March}, {Marshall}, {Melchior},
  {Menanteau}, {Miquel}, {Ogando}, {Plazas}, {Sanchez}, {Schubnell},
  {Sevilla-Noarbe}, {Smith}, {Soares-Santos}, {Sobreira}, {Suchyta}, {Swanson},
  {Tarle}, {Walker}, \& {DES Collaboration}}]{Zhang2017}
{Zhang}, Y., {Miller}, C.~J., {Rooney}, P., {et~al.} 2019{\natexlab{c}},
  \mnras, 488, 1

\end{thebibliography}
\end{document}